\documentclass[]{aa}
\usepackage[varg]{txfonts}
\usepackage{graphicx}
\usepackage{natbib}
\bibpunct{(}{)}{;}{a}{}{,} 
\graphicspath{{images/}}

\title{The GAPS Programme with HARPS-N at TNG}
\subtitle{XVIII. Two new giant planets around the metal-poor stars \\ HD 220197 and HD 233832
	    \thanks{Based on observations made with the HARPS-N spectrograph on the Italian Telescopio Nazionale Galileo (TNG) 
		    operated on the island of La Palma (Spain) by the INAF -- Fundación Galileo Galilei (Spanish Observatorio del 
		    Roque de los Muchachos of the Instituto de Astrofísica de Canarias) }
	  }
      
\author{D.~Barbato 		\inst{\ref{unito},\ref{oato}}
	\and A.~Sozzetti	\inst{\ref{oato}}
	\and K.~Biazzo		\inst{\ref{oact}}
	\and L.~Malavolta	\inst{\ref{unipd},\ref{oapd}}
	\and N.C.~Santos	\inst{\ref{ia-porto},\ref{dfa-up}}
	\and M.~Damasso		\inst{\ref{oato}}
	\and A.F.~Lanza		\inst{\ref{oact}}
	\and M.~Pinamonti	\inst{\ref{oato}}
	\and L.~Affer		\inst{\ref{oapa}}
	\and S.~Benatti		\inst{\ref{oapd}}
	\and A.~Bignamini	\inst{\ref{oats}}
	\and A.S.~Bonomo	\inst{\ref{oato}}
	\and F.~Borsa		\inst{\ref{oabrera}}
	\and I.~Carleo		\inst{\ref{oapd},\ref{unipd}}
	\and R.~Claudi		\inst{\ref{oapd}}
	\and R.~Cosentino	\inst{\ref{fgg}}
	\and E.~Covino		\inst{\ref{oana}}
	\and S.~Desidera	\inst{\ref{oapd}}
	\and M.~Esposito	\inst{\ref{oana},\ref{thuringer}}
	\and P.~Giacobbe	\inst{\ref{oato}}
	\and E.~Gonz\'{a}lez-\'{A}lvarez	\inst{\ref{oapa}}
	\and R.~Gratton		\inst{\ref{oapd}}
	\and A.~Harutyunyan	\inst{\ref{fgg}}
	\and G.~Leto		\inst{\ref{oact}}
	\and A.~Maggio		\inst{\ref{oapa}}
	\and J.~Maldonado	\inst{\ref{oapa}}
	\and L.~Mancini		\inst{\ref{unirm},\ref{mpia},\ref{oato}}
	\and S.~Masiero		\inst{\ref{oapa}}
	\and G.~Micela		\inst{\ref{oapa}}
	\and E.~Molinari	\inst{\ref{fgg},\ref{oaca}}
	\and V.~Nascimbeni	\inst{\ref{unipd},\ref{oapd}}
	\and I.~Pagano		\inst{\ref{oact}}
	\and G.~Piotto		\inst{\ref{unipd},\ref{oapd}}
	\and E.~Poretti		\inst{\ref{fgg},\ref{oabrera}}
	\and M.~Rainer		\inst{\ref{oabrera},\ref{oafi}}
	\and G.~Scandariato	\inst{\ref{oact}}
	\and R.~Smareglia	\inst{\ref{oats}}
	\and L.S.~Colombo	\inst{\ref{unipd}}
	\and L.~Di Fabrizio	\inst{\ref{fgg}}
	\and J.P.~Faria		\inst{\ref{ia-porto},\ref{dfa-up}}
	\and A.~Martinez~Fiorenzano	\inst{\ref{fgg}}
	\and M.~Molinaro	\inst{\ref{oats}}
	\and M.~Pedani		\inst{\ref{fgg}}
	}
\institute{Dipartimento di Fisica, Universit\`{a} degli Studi di Torino, via Pietro Giuria 1, I-10125 Torino, Italy \label{unito}
      \and INAF – Osservatorio Astrofisico di Torino, Via Osservatorio 20, I-10025 Pino Torinese, Italy \label{oato}
      \and INAF – Osservatorio Astrofisico di Catania, Via S. Sofia 78, I-95123, Catania, Italy \label{oact}
      \and Dipartimento di Fisica e Astronomia "G. Galilei", Universit\`{a} di Padova, Vicolo dell'Osservatorio 3, I-35122 Padova, 
	    Italy \label{unipd}
      \and INAF – Osservatorio Astronomico di Padova, Vicolo dell’Osservatorio 5, I-35122, Padova, Italy \label{oapd}
      \and Instituto de Astrof\'{i}sica e Ci\^{e}ncias do Espa\c{c}o, Universidade do Porto, CAUP, Rua das Estrelas, 4150-762 Porto, 
	    Portugal \label{ia-porto}
      \and Departamento de F\'{i}sica e Astronomia, Faculdade de Ci\^{e}ncias, Universidade do Porto, Rua do Campo Alegre, 4169-007 
	    Porto, Portugal \label{dfa-up}
      \and INAF – Osservatorio Astronomico di Palermo, Piazza del Parlamento 1, I-90134, Palermo, Italy \label{oapa}
      \and INAF – Osservatorio Astronomico di Trieste, via Tiepolo 11, I-34143 Trieste, Italy \label{oats}
      \and INAF – Osservatorio Astronomico di Brera, Via E. Bianchi 46, I-23807 Merate (LC), Italy \label{oabrera}
      \and Fundación Galileo Galilei - INAF, Rambla José Ana Fernandez Pérez 7, E-38712 Breña Baja, TF, Spain \label{fgg}
      \and INAF – Osservatorio Astronomico di Capodimonte, Salita Moiariello 16, I-80131, Napoli, Italy \label{oana}
      \and Thüringer Landessternwarte Tautenburg, Sternwarte 5, D-07778 Tautenburg, Germany \label{thuringer}
      \and Dipartimento di Fisica, Università di Roma Tor Vergata, Via della Ricerca Scientifica 1, 00133 Roma, Italy \label{unirm}
      \and Max Planck Institute for Astronomy, Königstuhl 17, 69117 Heidelberg, Germany \label{mpia}
      \and INAF – Osservatorio Astronomico di Cagliari, Via della Scienza 5, I-09047 Selargius (CA) , Italy \label{oaca}
      \and INAF – Osservatorio Astrofisico di Arcetri, Largo E. Fermi 5, I-50125 Firenze, Italy \label{oafi}
	}
	
\date{Received <date> / Accepted <date>}
\abstract{Statistical studies of exoplanets have shown that giant planets are more commonly hosted by metal-rich dwarf stars than 
	  low-metallicity ones, while such a correlation is not evident for lower-mass planets. The search for giant planets 
	  around metal-poor stars and the estimate of their occurrence $f_p$ is an important element in providing support to 
	  models of planet formation.}
	  {We present results from the HARPS-N search for giant planets orbiting metal-poor ($-1.0\leq\textnormal{[Fe/H]}\leq-0.5$ dex) 
	  stars in the northern hemisphere complementing a previous HARPS survey on southern stars in order to update the estimate of 
	  $f_p$.}
	  {High-precision HARPS-N observations of 42 metal-poor stars are used to search for planetary signals to be 
	  fitted using differential evolution MCMC single-Keplerian models. We then join our detections to the results of the previous 
	  HARPS survey on 88 metal-poor stars to provide a preliminar estimate of the two-hemisphere $f_p$.}
	  {We report the detection of two new giant planets around HD 220197 and HD 233832. The first companion has  
	  M$\sin{i}=0.20_{-0.04}^{+0.07}$ M$_{\rm Jup}$ and orbital period of $1728_{-80}^{+162}$ days, and for the second 
	  companion we find 
	  two solutions of equal statistical weight having periods $2058_{-40}^{+47}$ and $4047_{-117}^{+91}$ days 
	  and minimum masses of $1.78_{-0.06}^{+0.08}$ and $2.72_{-0.23}^{+0.23}$ M$_{\rm Jup}$, respectively. Joining our two 
	  detections with the three from the southern survey we obtain a preliminary and conservative estimate of global frequency of 
	  $f_p=3.84_{-1.06}^{+2.45}\%$ for giant planets around metal-poor stars.}
	  {The two new giant planets orbit dwarf stars at the metal-rich end of the HARPS-N metal-poor sample, corroborating previous 
	  results suggesting that giant planet frequency still is a rising function of host star [Fe/H]. We also note that all 
	  detections in the overall sample are giant long-period planets.}
\keywords{techniques: radial velocities - methods: data analysis - planetary systems - stars: abundances - stars: individual: HD 220197, HD 233832}

\begin{document}  
  \maketitle
  
  \section{Introduction} \label{sec:introduction}
    Amongst the physical properties of planet-host stars, mass and metallicity seem to have the biggest impact in promoting the 
    formation of giant (M$>20$ M$_\oplus$) planets; many studies have long shown that M dwarfs with $M_*<0.5$ M$_\odot$ are 
    less likely to host a Jupiter-like planet than Sun-like or more massive F and G main-sequence stars 
    \citep{butler2004,johnson2007,johnson2010,bonfils2013}, and that metal-rich stars have a much higher probability to be 
    orbited by at least one giant planet than lower metallicity stars 
    \citep{gonzalez1997,fischer2005,sozzetti2004,santos2001,santos2004,sozzetti2009,mortier2012}.
    \par Of special interest in recent years is the positive correlation between stellar metallicity and occurrence of giant planets,
    especially since such a correlation is not 
    found between host star metallicity and frequency of sub-Neptunian (R<4 R$_\oplus$, M$\sin{i}$<10 M$_\oplus$) planets 
    \citep{udry2006,sousa2008,sousa2011,mayor2011,buchhave2012,courcol2016}. In particular, \cite{mortier2012} reports that the fraction of stars hosting a 
    giant planet rises from 5\% for solar metallicity values to 25\% for metallicities that are twice that of the Sun; these 
    conclusions are also supported by results from the Kepler mission \citep[see e.g.][]{buchhave2018}.
    \par The correlation between host 
    star metallicity [Fe/H] and frequency of giant planets is 
    usually seen as strong evidence favouring core-accretion over disk instability formation models for giant planets. In the 
    core-accretion model \citep{pollack1996,ida2004,mordasini2009a,mordasini2012} giant planets are formed from the accretion of 
    material into solid cores until they are massive enough ($\sim$10 M$_\oplus$) to trigger a rapid agglomeration of gas, a process 
    more efficient in metal-rich disks. In the disk instability model \citep{boss1997,mayer2002,boss2002,boss2006} giant planets form 
    directly from the collapse of self-gravitating clumps of gas after the disruption of the proto-planetary disk, without requiring 
    the presence of solid cores.
    \par Recent works also show interesting correlations between metallicity regimes and the 
    class of giant planets found around the host star, stressing the importance of stellar metallicity as a proxy for protoplanetary 
    disk chemical composition and its role in driving planet formation and dynamical evolution. Metal-poor stars seem to host 
    planets that are more massive and with longer periods than those hosted by metal-rich stars. \cite{sozzetti2004} 
    argues for an anti-correlation between orbital period and host star metallicity, a result more recently supported by 
    \cite{mulders2016} based on an analysis of Kepler candidates; \cite{santos2017} also report that stars with planets more 
    massive than 4 $M_{Jup}$ are on average more metal-poor than stars hosting less massive planets.
    \par Similarily, \cite{maldonado2015b} report 
    that stars hosting hot Jupiters (defined as giant planets with semimajor axis $a<0.1$ AU) tend to have slightly 
    higher metallicities than stars orbited by more distant giants, further noticing that no hot Jupiter are found around stars having 
    metallicities lower than $-$0.6 dex. This result is confirmed by 
    the analysis of 59 cool Jupiter hosts and 29 hot Jupiter hosts in \cite{maldonado2018}, finding a significant deficit of hot Jupiter 
    planet hosts below $+$0.2 dex compared to cool Jupiter planet hosts and interpreting the different chemical characteristics of the 
    host stars of these planetary classes as distinct planetary populations with different evolutionary history; the same study also 
    notes that metal-poor stars hosting cool Jupiters have higher $\alpha$-element abundances than those hosting hot Jupiters, 
    suggesting that in metal-poor protoplanetary disks an overabundance of elements such as Mg, Si, Ti may compensate for the 
    lack of Fe in allowing the formation of giant planets according to the core-accretion model. Previous studies 
    \citep{haywood2008,haywood2009,kang2011,adibekyan2012a,adibekyan2012c} had similarly noted that planet-hosting stars having low [Fe/H] tend to be 
    enhanced in $\alpha$-elements.
    \par \cite{buchhave2018} also find 
    that stars hosting Jupiter analogues have average metallicities close to that of the Sun, while hot Jupiters as well as cool 
    eccentric ones are found around stars having higher metallicities, suggesting that planet-planet scattering mechanisms producing 
    more eccentric orbits are more common in metallic protoplanetary environments.    
    \par The correlation between stellar [Fe/H] and occurrence of giant planets has inspired the search for such planetary bodies 
    around stars specifically selected for their low [Fe/H] values, especially to determine the metallicity limit below 
    which no giant companions are formed. In 2003 a three-years survey using HIRES on the Keck I telescope was started 
    \citep[see][]{sozzetti2006,sozzetti2009}, observing $\sim$200 metal-poor ($-2\leq\textnormal{[Fe/H]}\leq-0.6$ dex) stars obtaining a 
    $\sim$10 ms$^{-1}$ 
    Doppler precision and finding no candidate planet. The null detection was therefore used to provide a 1-$\sigma$ upper limit of 
    0.67\% for the frequency of massive planets orbiting low-metallicity stars at orbital periods P$<3$ yr.    
    \par Also, one of the High Accuracy Radial velocity Planet Searcher \citep[HARPS, see][]{mayor2003} guaranteed time observations 
    (GTO) sub-samples was 
    explicitly build to search for giant planets orbiting metal-poor stars in the southern hemisphere, using the high precision of the  
    spectrograph ($\sim$1 ms$^{-1}$) to evaluate their frequency and low-metallicity formation 
    limit. The study of this 88 metal-poor stars sub-sample has found a total of three giant planets 
    (HD 171028 b, HD 181720 b, HD 190984 b) having minimum masses of 1.98, 0.37 and 3.1 M$_{jup}$ and orbital periods of 550, 956 and 
    4885 days, respectively. In addition to these detection, one yet unconfirmed planet was proposed around star HD 107094, 
    having minimum mass of 4.5 M$_{jup}$ and period of 1870 days \citep[see][]{santos2007,santos2010b,santos2011}.
    From these 3 giant planets over 88 target stars, \cite{santos2011} derive a frequency of Jupiter-mass planets around 
    metal-poor stars of $f_p=3.4_{-1.0}^{+3.2}\%$, a value that rises to $11.3_{-5.3}^{+4.9}\%$ when considering only the 34 stars in 
    the metallicity range in which the three detected planets have been found and with more than 3 measurements 
    ([Fe/H] between $-$0.40 and $-$0.60 dex), 
    and also that the null detection on the 32 sample stars with [Fe/H]$<-0.60$ dex and at least 6 datapoints implies a frequency of 
    $f_p<5\%$ for this subsample. All of these results are presented as conservative estimates, due to the possibility of having not 
    detected 
    existing giant companions on short periods due to unoptimal sampling, low number of measurements and considering a possible 
    fourth detection around HD 107094 and several linear trends in the analyzed sample. Interestingly, \cite{santos2011} further 
    note that the four stars around which the three planets and one candidate were 
    observed have [Fe/H] values ($-$0.48, $-$0.53, $-$0.49 and $-$0.51 dex) on the high-metallicity end of the sample, suggesting 
    that even for metal-poor stars the giant planet frequency is a rising function of the host star metallicity.
    \par A follow-up analysis of the metal-poor samples studied in \cite{sozzetti2009} and \cite{santos2011} was presented in 
    \cite{mortier2012}, showing that while no hot Jupiters are found, and are therefore rare around 
    such metal-poor stars($f_p<1\%$), the fraction of long-period giant planets is much higher, rising from $f_p<2.35\%$ at 
    [Fe/H]$\leq$-0.7 dex to $f_p=4.48_{-1.38}^{+4.04}\%$ for stars with [Fe/H]>-0.7 dex.
    \par Furthermore, \cite{johnson2012} find that the critical value for [Fe/H] under which giant planets are not formed is a 
    function of planetary distance $r$ from the host star, estimating a lower limit for this value of 
    [Fe/H]$_{crit}\simeq-1.5+\log{r}$.
    Interestingly, claims of 
    giant planets found around exceptionally metal-poor ([Fe/H]$\sim-2.0$ dex) stars were disproven by follow-up studies, such as the 
    case of the planetary systems proposed around the stars HIP 11952 \citep[see][]{desidera2013,muller2013} and HIP 13044
    \citep[see][]{jones2014}. We also note that that, as of the time of writing, the dwarf star with the lowest metallicity known 
    to host any giant planet is HD 155358 with [Fe/H]=$-0.62$ dex, hosting two planets with minimum masses of 0.99 and 0.82 
    M$_{jup}$ at periods of 194.3 and 391.9 days \citep[see][]{robertson2012,santos2013}
    \par In this paper we present results from the survey conducted with the High Accuracy Radial velocity Planet Searcher in the 
    Northern hemisphere \citep[HARPS-N, see][]{cosentino2012} at the Telescopio Nazionale Galileo (TNG) in La Palma within the the 
    observational programme Global 
    Architecture of Planetary Systems \citep[GAPS, see][]{covino2013,desidera2013} on a northern metal-poor sample. We report the 
    detection of two giant planets around stars HD 220197 and HD 233832 and a preliminary revision of the giant planet frequency 
    $f_p$.
    In Sect. 
    \ref{sec:north-sample} we describe the selection and observations conducted with HARPS-N on our northern sample. We 
    characterize the host stars HD 220197 and HD 233832 in Sect. \ref{sec:new-results-stars} before presenting our orbital solutions 
    in Sect. \ref{sec:new-results-planets} and the update of planetary frequency in Sect. \ref{sec:frequency}; finally we conclude 
    and discuss the overall results in Sect. \ref{sec:conclusions}.
    
  \section{The HARPS-N metal-poor sample} \label{sec:north-sample}
  
      \begin{figure}
	\includegraphics[width=\linewidth]{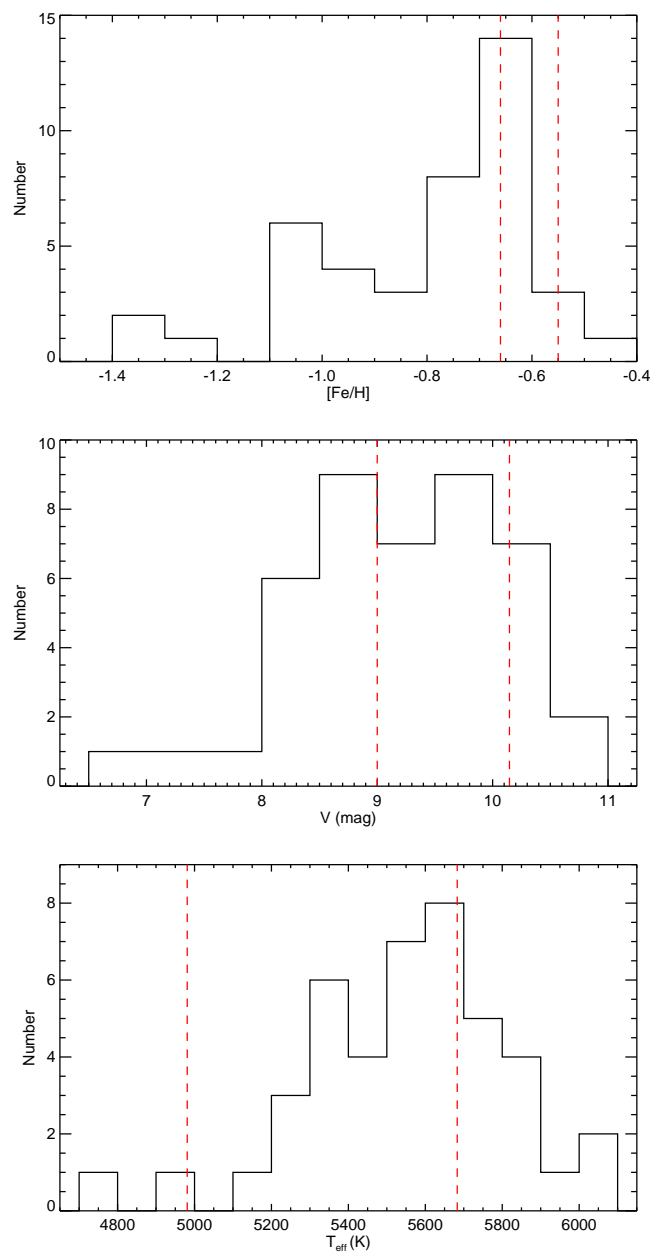}
	  \caption{Metallicity, magnitude and effective temperature distribution of the HARPS-N metal-poor sample discussed in this 
		    work. The values of stars HD 220197 and HD 233832 are indicated by the vertical dashed red lines.}
	  \label{fig:hist-gaps}     
      \end{figure}
    
	\begin{figure}
	  \includegraphics[width=\linewidth]{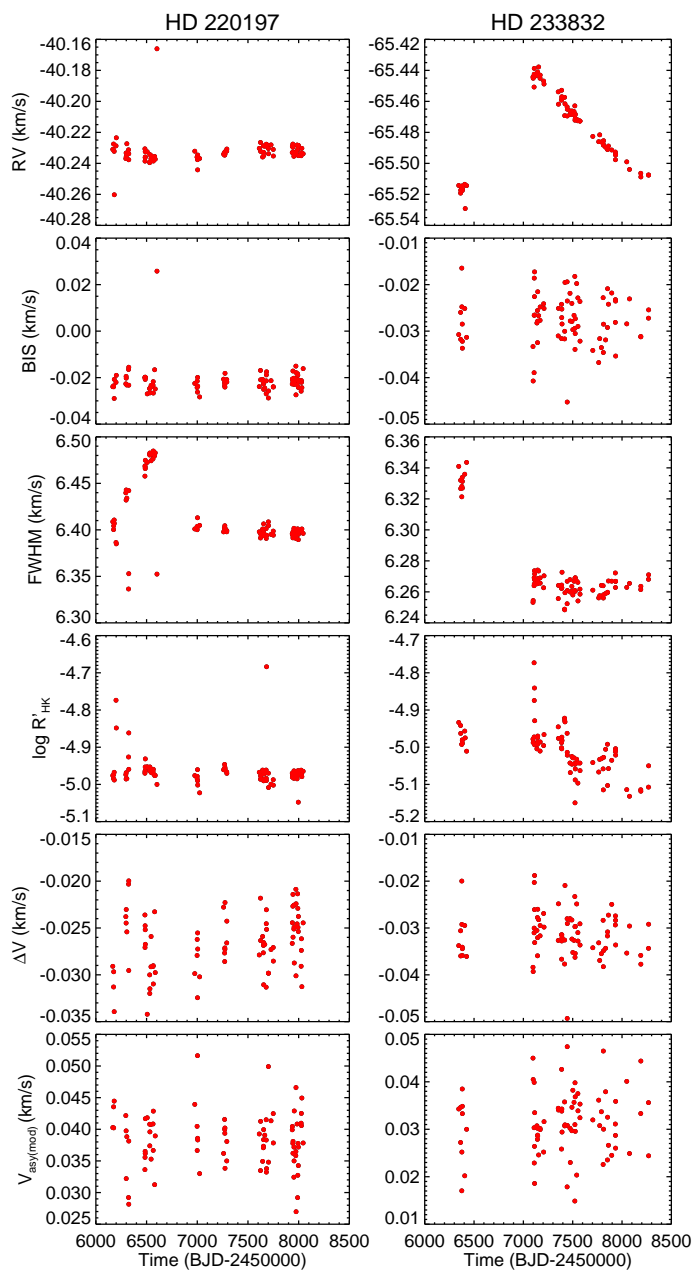}
	  \caption{Timeseries of measurements obtained with HARPS-N on HD 220197 (left column) and HD 233832 (right column). 
		    Rows top to bottom show radial velocities, 
		    Bisector Inverse Span, CCF FWHM, $\log R^{\prime}_{\rm HK}$, $\Delta$V and V$_{asy(mod)}$ respectively.}
	  \label{fig:timeseries-plot}
	\end{figure}
	
    The selected stars 
    were drawn from the sample of $\sim$200 stars previously observed with Keck/HIRES in a similar search for giant planetary 
    companions around metal-poor ($-2.0\leq\textnormal{[Fe/H]}\leq-0.6$ dex) stars \citep[see][]{sozzetti2006,sozzetti2009} in which 
    the targets selected from the Carney-Latham and Ryan samples of metal-poor, high-velocity field stars 
    \citep[see][]{carney1996,ryan1989,ryan1991} were surveyed 
    with a $\sim$10 ms$^{-1}$ precision. A sub-sample of 42 stars was therefore selected for observation with HARPS-N, having 
    metallicity approximatively between $-$1.0 and $-$0.5 dex, magnitude V$<11.0$ mag so to allow for photon noise precision of $\sim$1 ms$^{-1}$ 
    and being chromospherically quiet. The metallicity [Fe/H], magnitude $V$ and effective temperature $T_{eff}$ distributions of the 
    sample are shown in Fig. \ref{fig:hist-gaps}.      
    \par The 42-stars sample thus selected was monitored with HARPS-N from August 2012 to August 2018, obtaining a total of 1496 
    datapoints (a mean of 35 observations per star) with a mean error of 1.25 ms$^{-1}$; mean exposure time was 800 seconds and mean 
    signal-to-noise ratio (S/N) was 89.30.
    \par The high precision timeseries we obtained were searched for significant (false alarme probability FAP$\leq1\%$) 
    signals via a generalized Lomb-Scargle periodogram using the IDL routine \texttt{GLS} \citep{zechmeister2009}; in case of 
    detection of significant signals a single-planet orbital solution is tried out.
    \par Two stars in our sample (HD 220197, HD 233832) show a significant periodogram peak for which a successful search for 
    a planetary solution was made (see Sect. \ref{sec:new-results-planets}) and will therefore be the focus of this work's analysis. 
    The complete list of measurements collected 
    for these two stars is listed in Table \ref{table:rv-table} at the end of this paper.
    In Fig. \ref{fig:timeseries-plot} we show as red circles the timeseries of the radial velocities collected for HD 220197 and 
    HD 233832 and activity indexes bisector inverse span (BIS), full width at half maximum (FWHM) of the CCF,  
    $\log R^{\prime}_{\rm HK}$, $\Delta$V \citep[see][]{nardetto2006} and the new indicator $V_{asy(mod)}$ defined in 
    \cite{lanza2018} to avoid the known spurious dependencies with radial velocity variations of the indicator $V_{asy}$ defined 
    in \cite{figueira2013} and \cite{figueira2015}.
    In the 
    FWHM datasets the effect of a defocusing affecting the first portion of the HARPS-N observations is evident as a trend in the 
    datapoints, specifically affecting the first 31 datapoints collected for HD 220197 and the first 11 for HD 233832. This 
    instrumental defocusing was corrected in March 2014 and is absent from the following data. To correct for this 
    defocusing and improve the quality of the collected data by removing spurious correlations between radial velocities and FWHM 
    introduced by this instrumental effect, we follow \cite{benatti2017} in which the same effect 
    was encountered while characterizing the planetary system orbiting HD 108874 with HARPS-N data within the same time period. 
    After removing two spurious 
    observations at epoch $2456181.62$ and $2456602.56$ for HD 220197 
    (see Sect. \ref{sec:new-results-planets} for details), we performed a polynomial fit on the FWHM data affected by the defocusing
    and considered the best-fit residuals as corrected FWHM. We report the values of the corrected FWHM in parentheses in the fifth 
    column of Table \ref{table:rv-table} alongside the uncorrected FWHM values. This correction 
    successfully lowers the correlation between radial velocities and FWHM; for HD 220197 we find a Spearman correlation $r$ 
    between these two quantities of $-0.578$ before correction and of $-0.277$ after correction, while for HD 233832 $r$ varies from 
    $-0.2466$ to $0.008$ after correcting for the defocusing. In all following analysis we have used the corrected timeseries.
  
  \section{Stellar properties: HD 220197 and HD 233832} \label{sec:new-results-stars}
  
      \begin{table}
	    \caption{Stellar properties.}\label{table:stars-literature}
	      \centering
	      \begin{tabular}{l c c}
		\hline\hline
		  Parameter					& HD 220197		& HD 233832	\\
		\hline
		  $\alpha$ (J2000)				& $23^h 21^m 58.2^s$	& $11^h 26^m 05.5^s$	\\[3pt]
		  $\delta$ (J2000)				& $+16\degr37^{\prime}57^{\prime\prime}$	& $+50\degr22^{\prime}32^{\prime\prime}$	\\[3pt]
		  $\pi$ (mas)\tablefootmark{a}			& $15.496\pm0.046$	& $16.995\pm0.075$	\\[3pt]
		  $\mu_\alpha$ (mas yr$^{-1}$)\tablefootmark{a}	& $407.247\pm0.070$	& $-473.959\pm0.074$	\\[3pt]
		  $\mu_\delta$ (mas yr$^{-1}$)\tablefootmark{a}	& $-48.264\pm0.051$	& $124.167\pm0.087$	\\[3pt]
		  B (mag)\tablefootmark{b}			& $9.60\pm0.01$		& $10.92\pm0.04$	\\[3pt]
		  V (mag)\tablefootmark{b}			& $9.00\pm0.01$		& $10.146\pm0.060$	\\[3pt]
		  R (mag)\tablefootmark{c}			& $8.764\pm0.001$	& $9.912\pm0.001$	\\[3pt]
		  I (mag)\tablefootmark{c}			& $8.763\pm0.001$	& $9.047\pm0.040$	\\[3pt]
		  G (mag)\tablefootmark{a}			& $8.7448\pm0.0003$	& $9.899\pm0.004$	\\[3pt]
		  J (mag)\tablefootmark{b}			& $7.698\pm0.020$	& $8.544\pm0.024$	\\[3pt]
		  H (mag)\tablefootmark{d}			& $7.362\pm0.018$	& $8.042\pm0.061$	\\[3pt]
		  K (mag)\tablefootmark{b}			& $7.349\pm0.036$	& $8.013\pm0.020$	\\[3pt]
		\hline		  
	      \end{tabular}
	      \tablefoot{
			  \tablefoottext{a}{ retrieved from Gaia Data Release 2 \citep{gaia2018} }
			  \tablefoottext{b}{ retrieved from \cite{smart2014} }
			  \tablefoottext{c}{ retrieved from \cite{monet2003} }
			  \tablefoottext{d}{ retrieved from \cite{cutri2003} }
			}
      \end{table}
      
      \begin{table}
	  \caption{Spectroscopic stellar parameters as obtained using the two different analysis on equivalent width described 
		    in Sect. \ref{sec:new-results-stars}}	\label{table:star-spectroscopy}
	    \centering
	    \begin{tabular}{l c c}
	      \hline\hline
					& \multicolumn{2}{c}{HD 220197}	\\
	      \hline
		T$_{eff}$ 		& $5750\pm25$ 	& $5645\pm19$	\\[3pt]
		$\log{g}$ (cgs) 	& $4.40\pm0.13$ & $4.42\pm0.03$	\\[3pt]
		$\xi$ (kms$^{-1}$) 	& $1.17\pm0.05$ & $0.91\pm0.04$	\\[3pt]
		[Fe/H] 			& $-0.50\pm0.09$& $-0.55\pm0.02$\\[3pt]
	      \hline
					& \multicolumn{2}{c}{HD 233832}	\\
	      \hline
		T$_{eff}$ 		& $5075\pm75$ 	& $4961\pm35$ \\[3pt]
		$\log{g}$ (cgs) 	& $4.54\pm0.15$ & $4.48\pm0.07$ \\[3pt]
		$\xi$ (kms$^{-1}$) 	& $1.04\pm0.02$ & $0.19\pm0.31$ \\[3pt]
		[Fe/H] 			& $-0.54\pm0.09$& $-0.67\pm0.03$ \\[3pt]
	      \hline
	    \end{tabular}		 
      \end{table}
    Catalogue stellar 
    parameters for the two host stars are provided in Table \ref{table:stars-literature}, while the stellar parameters and elemental 
    abundances obtained from our spectroscopic analysis based on HARPS-N spectra are shown in Table \ref{table:stars-newderivation}.
    \par Effective temperature $T_{eff}$, surface gravity $\log{g}$, microturbulence velocity $\xi$ and iron abundance [Fe/H] were 
    measured through equivalent widths and using of a grid of Kurucz model atmospheres \citep{kurucz1993} and the spectral analysis 
    package MOOG \citep{sneden1973}. In particular, 
    $T_{eff}$ was derived by imposing that the \ion{Fe}{i} abundance does not depend on the excitation potential of the lines, $\xi$ by 
    imposing that the \ion{Fe}{i} abundance is independent on the line equivalent widths, and $\log{g}$ by the \ion{Fe}{i}/\ion{Fe}{ii} ionization 
    equilibrium condition. To account for possible differences in the calculation of equivalent widths we used 
    two different softwares, namely \texttt{IRAF} \citep{tody1993} and \texttt{ARES2} \citep{sousa2015}, to compute them from the 
    HARPS-N master spectra for HD 220197 and HD 233832 built from the coaddition of the individual spectra used for the radial 
    velocity measurements. The values of 
    $T_{eff}$, $\log{g}$, $\xi$ and [Fe/H] obtained from these two measurements of equivalent widths are respectively shown in the 
    first and second columns  of Table \ref{table:star-spectroscopy} and are generally in good agreement; in the following we use the 
    weighted mean of each parameter thus obtained, shown in Table \ref{table:stars-newderivation}.
    The differential elemental abundances with respect to the Sun were measured from our HARPS-N spectra following the method 
    detailed in \cite{damasso2015}, \cite{biazzo2015}, \cite{santos2013} and references therein. The first value of uncertainty on 
    elemental abundance is obtained from the measure of the equivalent width, while the second value is the root sum 
    square of the errors on abundance due to the uncertainties in stellar parameters $T_{eff}$, $\log{g}$ and $\xi$. 
    
      \begin{table}
	    \caption{Newly derived stellar parameters and elemental abundances. The first errors on elemental abundances 
		    refer to the measure of equivalent width, while the errors in parentheses are obtained from the root sum square 
		    of the abundance error caused by uncertainties on 
		    $T_{eff}$, $\log{g}$ and $\xi$.}\label{table:stars-newderivation}
	      \centering
	      \small
	      \begin{tabular}{l c c}
		\hline\hline
		  Parameter					& HD 220197		& HD 233832	\\
		\hline
		  Mass (M$_\odot$)\tablefootmark{a} 		& $0.91\pm0.02$		& $0.71\pm0.02$	\\[3pt]
		  Radius (R$_{\odot}$)\tablefootmark{a}		& $0.98\pm0.02$		& $0.68\pm0.03$	\\[3pt] 
		  Age (Gyr)\tablefootmark{b}			& $10.165\pm1.367$	& $5.417\pm4.165$	\\[3pt]
		  T$_{eff}$ (K)\tablefootmark{c}		& $5683\pm15$		& $4981\pm31$		\\[3pt]
		  $\log{g}$ (cgs)\tablefootmark{c}		& $4.42\pm0.03$		& $4.49\pm0.06$		\\[3pt]
		  $\xi$ (kms$^{-1}$)\tablefootmark{c}		& $1.01\pm0.03$		& $1.04\pm0.02$		\\[3pt]
		  $v_{macro}$ (kms$^{-1}$)			& $3.1$			& $1.8$			\\[3pt]
		  $v\sin{i}$ (kms$^{-1}$)			& $1.5\pm0.5$		& $0.8\pm0.5$		\\[3pt]
		  $\log R^{\prime}_{\rm HK}$			& $-4.96$		& $-5.01$		\\[3pt]
		  P$_{rot}$ (d)\tablefootmark{d}		& $\sim19$		& $\sim41$		\\[3pt]
		  [Fe/H]\tablefootmark{c}    & $-0.55\pm0.02$ ($+-0.06$) & $-0.66\pm0.03$ ($+-0.10$)  \\[3pt]
		  [\ion{C}{i}/H]   & $-0.26\pm0.09$ ($+-0.04$) & $+0.09\pm0.08$ ($+-0.07$) \\[3pt]
		  [\ion{Na}{i}/H]   & $-0.41\pm0.03$ ($+-0.02$) & $-0.60\pm0.04$ ($+-0.07$) \\[3pt]
		  [\ion{Mg}{i}/H]   & $-0.21\pm0.04$ ($+-0.02$) & $-0.42\pm0.04$ ($+-0.04$) \\[3pt]
		  [\ion{Al}{i}/H]   & $-0.30\pm0.17$ ($+-0.01$) & $-0.46\pm0.16$ ($+-0.05$) \\[3pt]
		  [\ion{Si}{i}/H]   & $-0.35\pm0.09$ ($+-0.01$) & $-0.55\pm0.08$ ($+-0.03$) \\[3pt]
		  [\ion{S}{i}/H]   & $-0.27\pm0.04$ ($+-0.04$) & $-0.27\pm0.09$ ($+-0.07$) \\[3pt]
		  [\ion{Ca}{i}/H]   & $-0.29\pm0.06$ ($+-0.03$) & $-0.51\pm0.12$ ($+-0.09$) \\[3pt]
		  [\ion{Ti}{i}/H]   & $-0.22\pm0.06$ ($+-0.03$) & $-0.45\pm0.12$ ($+-0.11$) \\[3pt]
		  [\ion{Ti}{ii}/H]   & $-0.40\pm0.09$ ($+-0.05$) & $-0.61\pm0.09$ ($+-0.06$) \\[3pt]
		  [\ion{Cr}{i}/H]   & $-0.49\pm0.06$ ($+-0.02$) & $-0.65\pm0.10$ ($+-0.08$) \\[3pt]
		  [\ion{Cr}{ii}/H]   & $-0.57\pm0.06$ ($+-0.05$) & $-0.66\pm0.10$ ($+-0.06$) \\[3pt]
		  [\ion{Ni}{i}/H]   & $-0.51\pm0.07$ ($+-0.02$) & $-0.73\pm0.09$ ($+-0.04$) \\[3pt]
		  [\ion{Zn}{i}/H]   & $-0.37\pm0.04$ ($+-0.02$) & $-0.61\pm0.06$ ($+-0.04$) \\[3pt]
		  [\ion{Y}{ii}/H]   & $-0.66\pm0.06$ ($+-0.05$) & $-0.88\pm0.10$ ($+-0.06$) \\[3pt]
		  [\ion{Zr}{ii}/H]   & $-0.51\pm0.03$ ($+-0.06$) & $-0.59\pm0.12$ ($+-0.06$) \\[3pt]
		  [\ion{Nd}{ii}/H]   & $-0.65\pm0.08$ ($+-0.06$) & $-0.62\pm0.18$ ($+-0.07$) \\[3pt]
		  [\ion{Cu}{i}/H]   & $-0.52\pm0.12$ ($+-0.03$) & $-0.73\pm0.16$ ($+-0.05$) \\[3pt]
		  [\ion{Eu}{ii}/H]   & $-0.44\pm0.07$ ($+-0.05$) & $-0.47\pm0.06$ ($+-0.07$) \\[3pt]
		  [\ion{La}{ii}/H]   & $-0.73\pm0.10$ ($+-0.06$) & $-0.82\pm0.08$ ($+-0.07$) \\[3pt]
		  [\ion{Mn}{i}/H]   & $-0.76\pm0.09$ ($+-0.03$) & $-0.80\pm0.09$ ($+-0.09$) \\[3pt]
		  [\ion{Ba}{ii}/H]   & $-0.73\pm0.10$ ($+-0.05$) & $-0.99\pm0.09$ ($+-0.05$) \\[3pt]
		\hline		  
	      \end{tabular}
	      \tablefoot{
			  \tablefoottext{a}{weighted mean between parameter calculated from PARAM 1.3 \citep[see][]{dasilva2006} 
					    and Yonsei-Yale isochrones \citep[see][]{yi2008}}
			  \tablefoottext{b}{calculated from PARAM 1.3}
			  \tablefoottext{c}{weighted mean between the values obtained from the two analyses on equivalent width 
					    described in Sect. \ref{sec:new-results-stars}}
			  \tablefoottext{d}{calculated following \cite{noyes1984} and \citet{mamajek2008}.}
			}
      \end{table}
    \par To compensate between model dependencies, we estimate the stellar mass and radius as the weighted mean of the values obtained 
    from the online tool PARAM 1.3 \citep[see][]{dasilva2006} and those obtained from the Yonsei-Yale isochrones \citep[see][]{yi2008}.      
    \par It can be noted that the photometric $T_{eff}$, [Fe/H] and M$_*$ reported in \cite{sozzetti2009} (5564 K, $-$0.65 dex and 
    0.83 M$_\odot$ for HD 220197; 4941 K, $-$0.74 dex and 0.69 M$_\odot$ for HD 233832) are lower than the spectroscopic values 
    obtained from HARPS-N, a result already noted in previous works 
    \citep[e.g.][]{biazzo2007,sozzetti2007,torres2012,tsantaki2013,maldonado2015a}.
    \par We synthesize spectral lines around 6200 \r{A} and 6700 \r{A} to obtain an estimate 
    of projected rotational velocity $v\sin{i}$ from fixed macroturbulence, instrument resolution and limb-darkening 
    coefficient, using the atmospheric models of \cite{kurucz1993}. 
    For HD 220197 assuming from the relations found in \cite{brewer2016} $v_{macro}=3.1$ kms$^{-1}$ we find a projected 
    rotational velocity of $v\sin{i}=1.5\pm0.5$ kms$^{-1}$, while for the cooler HD 233832 we assume from the same relations 
    $v_{macro}=1.8$ kms$^{-1}$ and obtain a projected rotational velocity of $0.8\pm0.5$ kms$^{-1}$, below the $\sim$2 kms$^{-1}$ 
    resolution of HARPS-N, 
    suggesting a very slow stellar rotation unless the star is observed nearly pole-on. 
    From this estimates of $v\sin{i}$ we can give upper 
    limits for the rotational period as $P_{rot}=2\pi R_* / v\sin{i}$, obtaining values of 31 days for HD 220197 and 43 
    for HD 233832.
    Having from our observations mean values of $\log R^{\prime}_{\rm HK}$ of $-$4.96 for 
    HD 220197 and $-$5.01 for HD 233832 (see Sect. \ref{sec:new-results-planets} and Fig. \ref{fig:timeseries-plot}) we can also 
    provide 
    analytical estimates of the rotation period $P_{rot}$ using the empirical relations from \cite{noyes1984} and \cite{mamajek2008}, 
    obtaining values of $\sim19$ days and $\sim41$ days for HD 220197 and HD 233832 respectively.
    \par We also note that the Second Data Release (DR2) of the astrometric satellite Gaia \citep[see][]{gaia2016,gaia2018} has 
    confirmed the existence of a cooler 
    (T$_{eff}$=$3721_{-70}^{+230}$ K) and fainter (G=$12.71$ mag) stellar companion for HD 233832 at comparable parallax of 
    $17.066\pm0.053$ mas and at 4.8 arcseconds of angular 
    separation, which at a distance of 59 pc (see Table \ref{table:stars-literature} for the parallax value of the star) implies a 
    projected separation of 280 AU. This stellar object was listed as a possible companion for HD 233832 in previous catalogs 
    \citep[e.g.][]{cutri2003} but in the absence of measurements of parallax 
    its was not possible to confirm its binarity nature before the release of Gaia DR2 astrometry measurements.
    Assuming a metallicity value similar to that of 
    the primary star we use the Yonsei-Yale isochrones to provide a first estimate of mass and radius for this companion star 
    respectively of $0.420\pm0.049$ M$_\odot$ and $0.367\pm0.026$ R$_\odot$. Having therefore an estimate of the mass and orbital 
    projected separation of both stellar components, we can give a first assessment of the order of magnitude of the stellar 
    companion's orbital period as P$_B\sim$4400 yr.
    \par
\begin{figure}
	  \includegraphics[width=\linewidth]{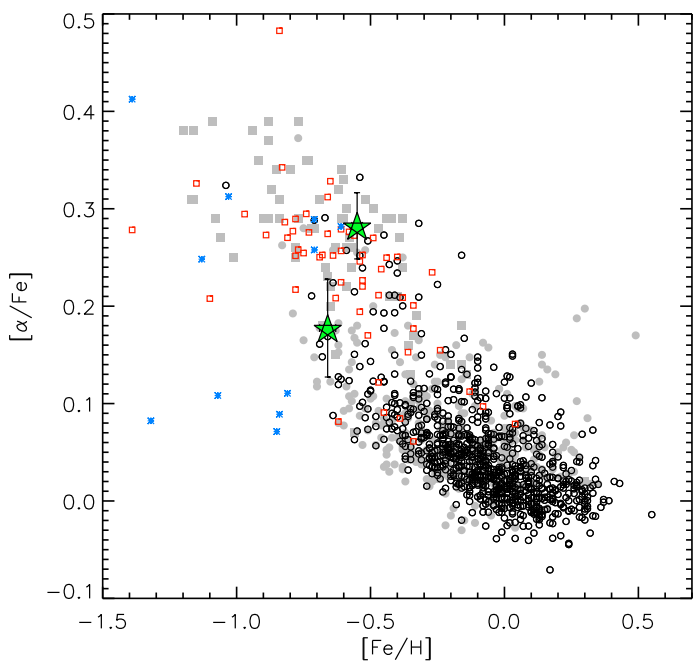}
	  \caption{[$\alpha$/Fe] versus [Fe/H] for HD 220197 (upper star symbol) and HD 233832 (lower star symbol). Thin-disk, 
		    thick-disk, and halo stars are shown with circles, squares, and asterisks, respectively 
		    (filled symbols: \cite{soubiran2005}; open symbols: \cite{adibekyan2012b}).
		  }
	  \label{fig:star-population}
	\end{figure}

 The results on elemental abundances allow us to investigate to which population the two targets belong. We thus considered 
    the abundances of field stars listed in the catalogs by \cite{soubiran2005} and \cite{adibekyan2012b} and applied the 
    prescriptions reported in \cite{biazzo2015}. We thus used as abundance of $\alpha$-elements that obtained from magnesium, 
    silicon, calcium, and titanium. Fig. \ref{fig:star-population} shows the position of HD 220197 and HD 233832 in the [$\alpha$/Fe] 
    versus [Fe/H] diagram. Based on these chemical indicators, the star HD 220197 seems to be more likely a thick-disk star than 
    HD 233832, the latter lying in a metallicity region populated by both thin and thick-disk stars. 
    \par Another way to classify the two stars as either thin- or thick-disk objects is to use statistical indicators purely based 
    on kinematics, that can then be compared with our inference based on the chemical indicator [$\alpha$/Fe]. In order to calculate 
    the likelihood of any given object belonging to either of the two populations on the basis of its Galactic kinematics, a number 
    of approaches can be adopted. We elect to carry out population assignments using the classifications by 
    \cite{bensby2003,bensby2005}. First, we combine systemic radial velocity and Gaia DR2 proper motion and parallax for both stars 
    to calculate the Galactic velocity vector (U, V, W, with U positive toward the Galactic anticenter) with respect to the local 
    standard of rest (LSR), adjusting for the standard solar motion ($U_\odot$, $V_\odot$, $W _\odot$) = ($-8.5$, $13.38$, $+6.49$) 
    kms$^{-1}$ (following Coskunoglu et al. 2011). Then, we calculate the thick disc–to–thin disk probability ratio $TD/D$ using the 
    prescriptions of \cite{bensby2003,bensby2005} for the velocity dispersion and asymmetric drift of the assumed Gaussian velocity 
    ellipsoid for the two populations, and the observed fractions of each population in the solar neighborhood (4\% and 96\%, 
    respectively). \cite{bensby2003} suggest as threshold to clearly identify thick- and thin-disk stars values of $TD/D\geq10$ and 
    $TD/D\leq0.1$, respectively. Finding for HD 233832 and HD 220197 $TD/D=1113$ and $TD/D=2.8$, respectively, HD 233832 is rather 
    clearly a thick-disk object, while for HD 220197 the evidence is for an object with kinematics intermediate 
    between that of thin and thick disk.
    \par Using the results from both methods, we can therefore reasonably classify both HD 220197 and HD 223832 as thick-disk object 
    within the uncertainties of the disk populations.
  
  \section{Radial velocity analysis} \label{sec:new-results-planets}
  
    Having found significant peaks in the radial velocities periodograms for stars HD 220197 and HD 233832, we searched for a 
    single-planet orbital solution via a differential evolution Markov chain Monte Carlo method \citep{eastman2013,desidera2004}, 
    the nine free parameters being inferior conjuction epoch $T_c$, orbital period $P$, 
    $\sqrt{e}\cos{\omega}$, $\sqrt{e}\sin{\omega}$, semiamplitude $K$, and 
    a zero-point radial velocity $\gamma$ and an uncorrelated jitter term $j$ for each instrument (Keck and HARPS-N). Uninformative 
    priors were used for all parameters. 
    Eighteen chains were run simultaneously and reached convergence and good mixing according to the criteria established 
    in \citep{eastman2013}.
    \par To ensure that the detected signals are not of stellar origin we search for correlations betweeen 
    radial velocities and activity indexes bisector inverse span (BIS), full width at half maximum (FWHM) of the CCF,  
    $\log R^{\prime}_{\rm HK}$, $\Delta$V and $V_{asy(mod)}$. 
    To obtain values of BIS, FWHM, $\Delta$V and $V_{asy(mod)}$ from our HARPS-N spectra 
    we use the \texttt{IDL} procedure presented in \cite{lanza2018}, while $\log R^{\prime}_{\rm HK}$ is obtained as detailed in 
    \cite{lovis2011}.
    \par The fitted and derived parameters and their 1$\sigma$ uncertainties, taken as the medians of the 
    posterior distributions and their 34.13\% intervals, are listed in Table \ref{table:fit-results} and discussed in the 
    following paragraphs.
      	
    \paragraph{HD 220197}

      We observed this star with HARPS-N from August 2012 to October 2017, obtaining 88 measurements with mean S/N of $\sim$110 
      that we join with the 5 Keck datapoints from \cite{sozzetti2009}; we note that the seven HARPS-N spectra obtained between 
      epochs $2456166.69$ and $2456201.60$ were taken with only the spectral orders falling on the blue side of the CCD. From these 
      datapoints we exclude the one taken at epoch $2456181.62$ due to low S/N and the one taken at epoch $2456602.56$, 
      which shows highly discrepant values in radial velocity, BIS and FWHM compared to the mean values of the timeseries 
      (see Fig. \ref{fig:timeseries-plot} and Table \ref{table:rv-table}) and therefore suggesting that this particular observation 
      is affected by an instrumental effect that we are not able to completely correct.
      \par
\begin{figure}
	  \includegraphics[width=\linewidth]{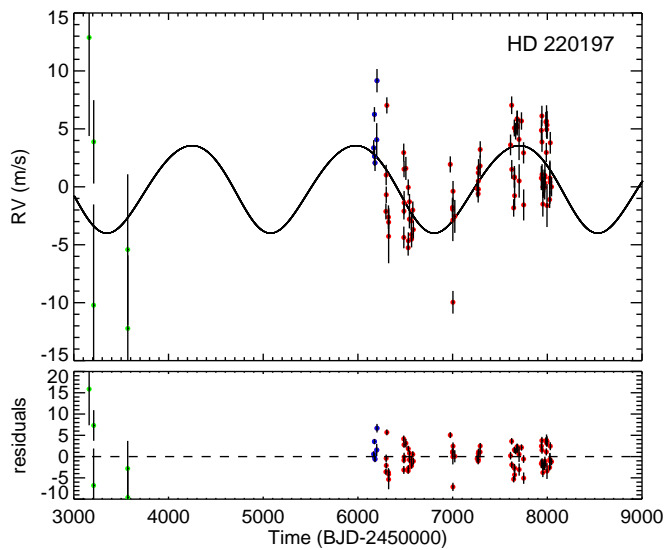}
	  \caption{Orbital fit for the planet HD 220197 b. In the top panel, our best-fit solution in shown as a black curve over the 
		    literature datapoints from Keck (green) and our HARPS-N observations (blue for blue-chip data, red for 
		    full-chip data). The bottom panel shows the residual radial velocities.}
	  \label{fig:hd220197-fit}
	\end{figure}

\begin{figure}
	  \includegraphics[width=\linewidth]{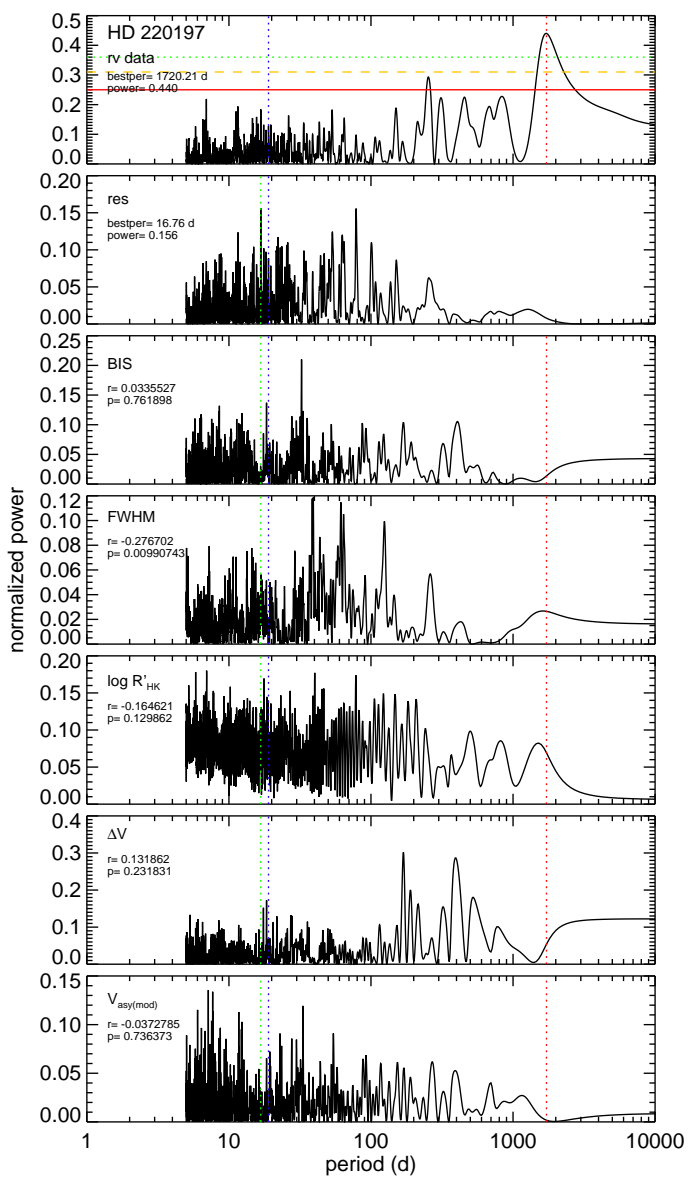}
	  \caption{Activity indexes periodograms for the star HD 220197. The two top panels show the periodograms for the radial 
		    velocity data and post-fit residuals; the most significant period value and power are shown in the upper right 
		    corner, while the horizontal lines indicate the false alarm probability levels of 10\% (solid red), 
		    1\% (dashed orange) and 0.1\% (dotted green). The following lower panels show the periodograms for bisector 
		    inverse span, FWHM, $\log R^{\prime}_{\rm HK}$, $\Delta$V and V$_{asy(mod)}$ with Spearman correlation rank $r$ 
		    and significance $p$ with radial velocity data shown in each panel. The most significant 
		    periods for radial velocity data and residuals are highlighted in all panels respectively as a red and 
		    green vertical dotted line, while the star's rotation period is indicated by a blue vertical dotted line.}
	  \label{fig:hd220197-activity}
	\end{figure}

 The periodogram of the timeseries (see top panel of Fig. \ref{fig:hd220197-activity}) shows a highly significant peak 
      at $\sim$1720 days with a false alarm probability of $0.01\%$ as calculated via bootstrap method which appears to be 
      uncorrelated with any of the activity indexes analyzed (see bottom panels of Fig. \ref{fig:hd220197-activity}) and therefore 
      not of clear stellar origin. The data is best fitted by a Keplerian curve (see Fig. \ref{fig:hd220197-fit} and 
      first column of Table \ref{table:fit-results}) having semiamplitude K=$3.78_{-0.72}^{+1.78}$ ms$^{-1}$, period 
      P=$1728_{-80}^{+162}$ days and eccentricity e=$0.187_{-0.132}^{+0.279}$, from which we obtain a planetary minimum mass of 
      $0.20_{-0.04}^{+0.07}$ M$_{\rm Jup}$ and semimajor axis of $2.729_{-0.085}^{+0.168}$ AU.
      \par To account for any offset or inconsistency between the half-chip and full-chip 
      data we also search for an orbital solution treating the two groups of data as independent datasets allowing for an offset 
      between them. The resulting solution features a Bayesian Information Criterion value (BIC) of 326.62, similar to the 
      BIC value of 321.69 obtained for the solution lacking any distinction between full-chip and half-chip data shown in Fig. 
      \ref{fig:hd220197-fit} and Table \ref{table:fit-results}. The introduction on an offset between half-chip and full-chip data is 
      not clearly statistically preferred and we therefore find no compelling reason to treat them as independent datasets.
      A similar case is represented by \cite{desidera2013}, in which the inclusion or lack of an offset between half-chip and full-chip 
      data collected with HARPS-N also turned out to be non significant in the analysis of the radial velocity data of the metal-poor 
      star HIP 11952.
      \par While a low-power peak can be found in both FWHM and $\log R^{\prime}_{\rm HK}$ periodograms (see fourth and fifth 
      panels of Fig. \ref{fig:hd220197-activity}) near the proposed orbital period for HD 220197 b, we stress that both are 
      non-significant peaks, having a FAP of 100\% and 99.4\% respectively.
      \par The post-fit residual data show a maximum peak around a period of $\sim$17 days, near our expected stellar rotation 
      period of 19 days (see Sect. \ref{sec:new-results-stars}), and another peak of comparable power at $\sim$80 days; 
      both peaks have an associated FAP of 55\% and are 
      therefore non-significant (see second panel of Fig. \ref{fig:hd220197-activity}).
    
    \paragraph{HD 233832}
	
      We monitored this star with HARPS-N from February 2013 to May 2018, obtaining 80 HARPS-N measurements with mean S/N of 
      $\sim$74; an additional 5 Keck datapoints were collected from \cite{sozzetti2009}.
      \par The HARPS-N data 
      clearly show a $\sim$80 ms$^{-1}$ variation in radial velocities, although we note that our observations failed to 
      satisfactorily 
      sample the minimum and rising portions of this variation. It may be argued that such a variation in radial velocity could be 
      related to the stellar companion recently confirmed by Gaia at angular separation of 4.8 arcseconds and for which we provide 
      a first estimate of M$_B$=$0.420\pm0.049$ M$_\odot$ and R$_B$=$0.367\pm0.026$ R$_\odot$ 
      (see Sect. \ref{sec:new-results-stars}). We however propose that this is not the case; following the example set in 
      \cite{torres1999} we can estimate the acceleration $d(RV)/dt$ caused by the stellar companion on the primary as:
	\begin{equation}	\label{eq:binary-dvdt}
	 \frac{d(RV)}{dt}= G \frac{M_B}{a^2 (1-e)} \frac{(1+\cos{v})\sin{(v+\omega)}\sin{i}}{(1+\cos{E})(1-e\cos{E})}
	\end{equation}
      being $a=a_A (M_A+M_B)/M_B$ the semimajor axis of the relative orbit, $v$ the true anomaly, $i$ the mutual inclination and $E$ 
      the eccentric anomaly. Having no estimate on the orbital elements of the 
      companion star except for the projected separation $\sim$280 AU, we generated $10^5$ possible combinations of orbital elements 
      $(e,v,\omega,E,i)$ from which we obtain a mean acceleration of $0.39$ ms$^{-1}$yr$^{-1}$; in addition to this we estimate the 
      maximum acceleration for a circular, edge-on stellar orbit to be $0.14$ ms$^{-1}$yr$^{-1}$. Both values, we note, are 
      exceptionally low and would cause a variation of at most $\sim$5.82 ms$^{-1}$ over our 15 years baseline. 
      We therefore argue that the origin of the observed $\sim$80 ms$^{-1}$ variation in the radial velocities of HD 233832 is not 
      the gravitational influence of its long-period stellar companion.	
      \par
\begin{figure}
	  \includegraphics[width=\linewidth]{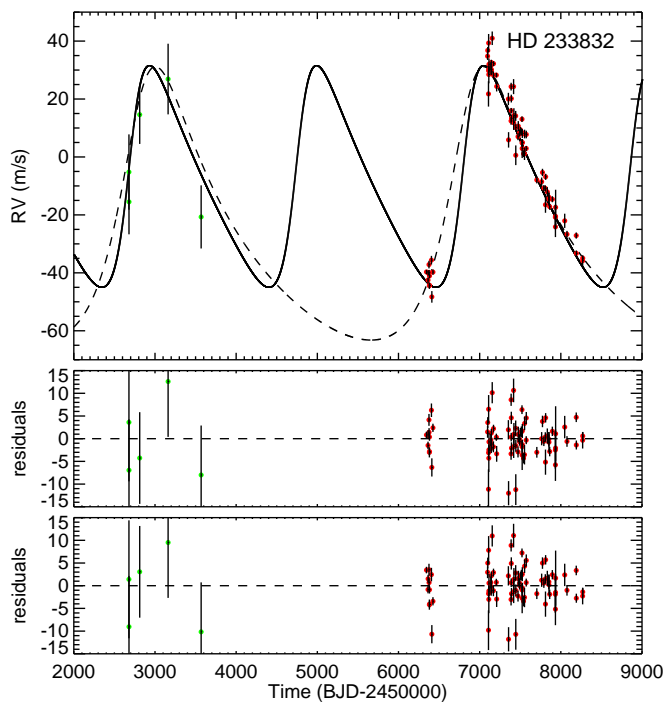}
	  \caption{Same as Fig. \ref{fig:hd220197-fit} but for the planet HD 233832 b. Literature Keck datapoints are show in green and our 
		    HARPS-N data are in red. The solid black curve shows the solution obtained setting prior on orbital period centered 
		    on 2000 days, while the dashed curve is derived from orbital prior centered around 4000 days. The second and third 
		    panels show respectively the residuals from the 2000 and 4000 days solutions.}
	  \label{fig:hd233832-fit}
	\end{figure}

\begin{figure}
	  \includegraphics[width=\linewidth]{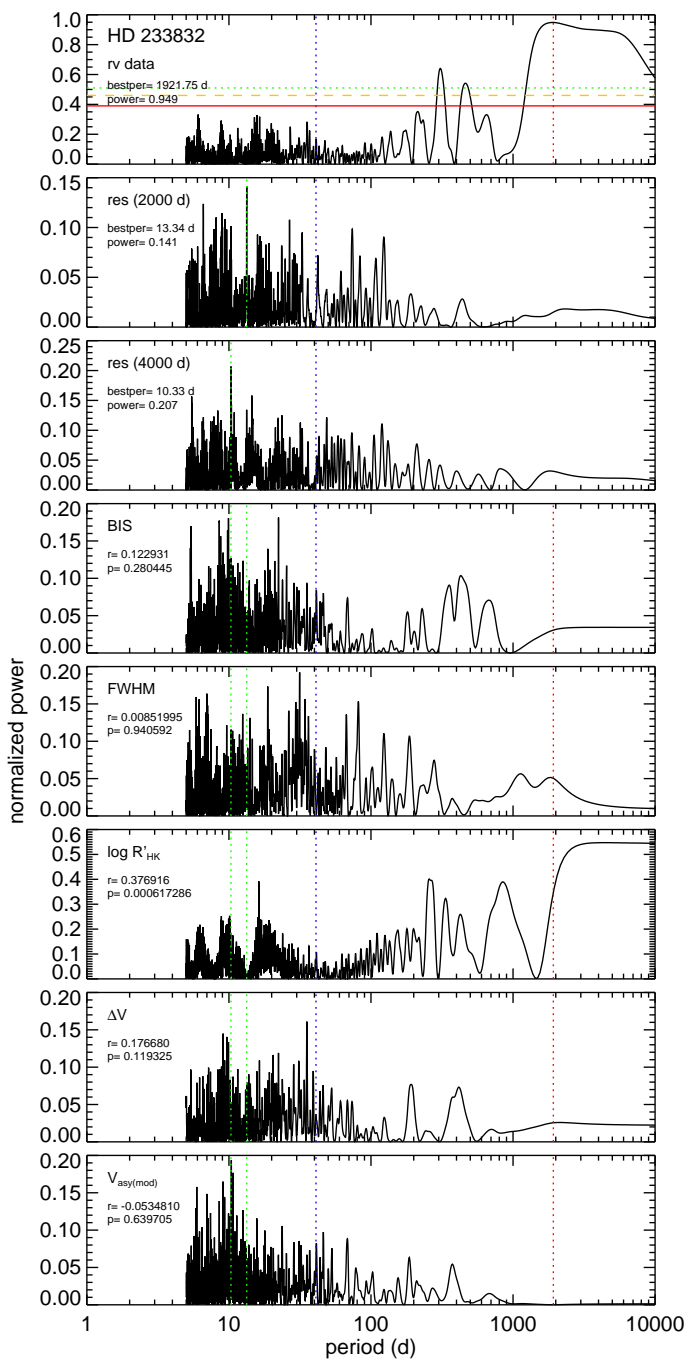}
	  \caption{Same as Fig. \ref{fig:hd220197-activity} but for the star HD 233832.}
	  \label{fig:hd233832-activity}
	\end{figure}

 The radial velocities periodogram (see top panel of Fig. \ref{fig:hd233832-activity}) shows a region of significant power 
      between 1000 and 6000 days, peaking around 1920 days with a FAP of $0.01\%$. The MCMC we launched with uniform priors returned 
      a Keplerian  best-fit with orbital period P=$2106_{-57}^{+813}$ days showing a large upper error bar; this 
      solution also has a period posterior distribution featuring two peaks of comparable likelihood around P$\sim$2000 d and 
      P$\sim$4000 d. To try and solve this apparent degeneracy in period, we 
      fit again the data setting Gaussian priors on the orbital period centered on 2000 and 4000 days with width of 500 days in order 
      to determine which period is statistically favourite; a comparison between the results of these fits is shown in 
      Fig. \ref{fig:hd233832-fit} and Table \ref{table:fit-results}.
	
      \par We thus find two solutions mainly distinguished by their values in semiamplitude, period and minimum planetary 
      mass. Choosing a prior centered around 2000 days returns a Keplerian solution having K=$38.29_{-1.35}^{+2.08}$ ms$^{-1}$, 
      P=$2058_{-40}^{+47}$ days and M$\sin{i}=1.78_{-0.06}^{+0.08}$ M$_{\rm Jup}$, while the prior around 4000 days returns a 
      solution having K=$47.18_{-3.21}^{+3.63}$ ms$^{-1}$, P=$4047_{-117}^{+91}$ days and M$\sin{i}=2.72_{-0.23}^{+0.23}$ 
      M$_{\rm Jup}$.
      It is clearly seen by comparing the Bayesian Information Criterion value for each solution (of 355.02 and 357.16 respectively)
      that none is strongly preferred above the other, although the $P\sim$2000 d solution is formally preferred for having a 
      slightly lower BIC value; no solid conclusion on the value of the planet's orbital period is therefore possible with the 
      available data.
      \par While the presence of the giant planet is clear, the values of its orbital period 
      and  mass remain ambiguous and more data with better sampling of the radial velocity minimum and its rise to 
      maximum are needed to discriminate between the two solutions. Another way to resolve this ambiguity may 
      come from observations with the astrometric satellite Gaia, which by the end of its five-years mission will have observed a
      significant portion of both proposed orbits. We can calculate the astrometric signatures produced on the host star by these 
      two possible solutions as:
	\begin{equation}	\label{eq:alpha}
	 \alpha=\frac{M_p}{M_*}\frac{a}{d}
	\end{equation}
      where $\alpha$ is in arcseconds, if planetary and stellar mass are given in solar mass units, semimajor axis $a$ in AU and 
      stellar distance $d$ in parsec. We then find $\alpha\sim115$ $\mu$as for the P$\sim$2000 d solution and 
      $\alpha\sim278$ $\mu$as for the P$\sim$4000 d solution, both expected to be detected at high S/N by Gaia for a star this 
      bright (G$=9.899$ mag).
      
	\begin{table*}
	  \caption{Orbital fit results}\label{table:fit-results}
	  \centering
	    \begin{tabular}{l c c c c c c}
	      \hline\hline
					& \multicolumn{2}{c}{HD 220197 b}			& \multicolumn{4}{c}{HD 233832 b}\\
		Parameter		& &							& \multicolumn{2}{c}{(prior P$\sim$2000 d)}		& \multicolumn{2}{c}{(prior P$\sim$4000 d)}\\
	      \hline
		K (ms$^{-1}$)		& \multicolumn{2}{c}{$3.78_{-0.72}^{+1.78}$}		&\multicolumn{2}{c}{$38.29_{-1.35}^{+2.08}$}		&\multicolumn{2}{c}{$47.18_{-3.21}^{+3.63}$}\\[3pt]
		P (days)		& \multicolumn{2}{c}{$1728_{-80}^{+162}$}		&\multicolumn{2}{c}{$2058_{-40}^{+47}$}			&\multicolumn{2}{c}{$4047_{-117}^{+91}$}\\[3pt]
		$\sqrt{e}\cos{\omega}$	& \multicolumn{2}{c}{$-0.227_{-0.386}^{+0.358}$}	&\multicolumn{2}{c}{$0.143_{-0.069}^{+0.074}$}		&\multicolumn{2}{c}{$0.481_{-0.075}^{+0.055}$}\\[3pt]
		$\sqrt{e}\sin{\omega}$	& \multicolumn{2}{c}{$0.127_{-0.251}^{+0.237}$}		&\multicolumn{2}{c}{$-0.577_{-0.044}^{+0.042}$}		&\multicolumn{2}{c}{$-0.384_{-0.058}^{+0.046}$}\\[3pt]
		$T_c$			& \multicolumn{2}{c}{$2456416.9_{-83.2}^{+138.3}$}	&\multicolumn{2}{c}{$2457688.8_{-32.5}^{+27.6}$}	&\multicolumn{2}{c}{$2457854.6_{-52.8}^{+52.8}$}\\[3pt]
		e			& \multicolumn{2}{c}{$0.187_{-0.132}^{+0.279}$}		&\multicolumn{2}{c}{$0.359_{-0.039}^{+0.046}$}		&\multicolumn{2}{c}{$0.381_{-0.028}^{+0.029}$}\\[3pt]
		$\omega$ (deg)		& \multicolumn{2}{c}{$159.011_{-80.914}^{+54.933}$}	&\multicolumn{2}{c}{$283.903_{-6.901}^{+7.543}$}	&\multicolumn{2}{c}{$321.365_{-8.792}^{+6.230}$}\\[3pt]
		M$\sin{i}$ (M$_J$)	& \multicolumn{2}{c}{$0.20_{-0.04}^{+0.07}$}		&\multicolumn{2}{c}{$1.78_{-0.06}^{+0.08}$}		&\multicolumn{2}{c}{$2.72_{-0.23}^{+0.23}$}\\[3pt]
		a (AU)			& \multicolumn{2}{c}{$2.729_{-0.085}^{+0.168}$}		&\multicolumn{2}{c}{$2.827_{-0.039}^{+0.045}$}		&\multicolumn{2}{c}{$4.438_{-0.090}^{+0.070}$}\\[3pt]
		T$_{peri}$ (days)	& \multicolumn{2}{c}{$2456636.3_{-428.0}^{+196.1}$}	&\multicolumn{2}{c}{$2456815.8_{-88.0}^{+82.6}$}	&\multicolumn{2}{c}{$2456874.9_{-89.9}^{+75.4}$}\\[3pt]
	      \hline
					& $\gamma$		     & jitter			& $\gamma$		     & jitter			& $\gamma$		     & jitter\\
					& (ms$^{-1}$)		     & (ms$^{-1}$)		& (ms$^{-1}$)		     & (ms$^{-1}$)		& (ms$^{-1}$)		     & (ms$^{-1}$)\\[3pt]
	      \hline
		Keck			& $2.22_{-6.05}^{+5.50}$     & $9.57_{-5.57}^{+9.65}$	& $-16.23_{-9.80}^{+10.64}$  & $10.55_{-7.59}^{+17.12}$ & $-38.78_{-10.10}^{+13.03}$ & $10.28_{-7.38}^{+17.31}$\\[3pt]
		HARPS-N			& $-0.92_{-0.58}^{+0.44}$    & $2.62_{-0.22}^{+0.25}$	& $-7.29_{-3.03}^{+3.53}$    & $3.18_{-0.33}^{+0.38}$ 	& $-28.84_{-1.82}^{+1.88}$   & $3.35_{-0.35}^{+0.39}$\\[3pt]
	      \hline
		BIC			& \multicolumn{2}{c}{321.69}				&\multicolumn{2}{c}{355.02}				&\multicolumn{2}{c}{357.16}\\[3pt]
	      \hline
	    \end{tabular}
	\end{table*}
	
      \par The periodograms on residual data for the two different orbital solutions (see the second and 
      third panels of Fig. \ref{fig:hd233832-activity}) show major peaks at around 13 days for the 
      P$\sim$2000 d solution and around 10 days for the P$\sim$4000 d solution with associated FAP of 
      27\% and 52\% respectively and therefore both non-significant.
      \par We also find in the periodogram of activity 
      index $\log R^{\prime}_{\rm HK}$ a clue of an uncorrelated long-period activity trend in the same period range of our proposed 
      planetary solutions (see sixth panel of Fig. \ref{fig:hd233832-activity}). While the blending of the target star spectra with 
      a nearby star can produce similar effects in the activity indexes \citep[see][]{santos2002}, the known star nearest to 
      HD 233832 is its companion confirmed by Gaia observations at 4.8 arcseconds, an angular distance much higher than the 1 
      arcsecond aperture on the sky of HARPS-N fibers and therefore unlikely to cause such an effect. While it may also be argued 
      that such an activity trend may in fact produce the observed 
      radial velocity variation of HD 233832 and therefore mimic the presence of a massive planet, it is worthy of note 
      that most cases of activity-induced radial velocity signals mimicking giant planets have semiamplitudes of 10 ms$^{-1}$ or less 
      \citep[see][for recent examples]{endl2016,johnson2016,carolo2014}, much lower than the $\sim$40 ms$^{-1}$ semiamplitude 
      observed in our HARPS-N data; the 11-year activity cycle of the Sun itself produces a radial velocity variation of 
      $4.98\pm1.44$ ms$^{-1}$, as detailed in \cite{lanza2016}. Also, the analysis on 304 FGK stars reported in \cite{lovis2011} and 
      conducted with HARPS concludes 
      that while 61\% of old solar-type stars have a detectable activity cycle inducing long-period radial velocity signals, the 
      semiamplitude of the effect is limited in the worst case of their sample to $\sim$11 ms$^{-1}$ and more typical values are 
      usually less than 3 ms$^{-1}$.
      \par To further support this point, we can give an estimate of the radial velocity perturbation caused by different activity 
      effects related to stellar rotation using the relations found in \cite{saar1997,saar1998,saar2000} between projected 
      rotational velocity $v\sin{i}$, macroturbolent velocity $v_{macro}$ and percentage of surface area covered by spot 
      distribution inhomogeneity $f_s$ (ranging from zero for old, inactive stars 
      to several percent for active stars); we also note that \cite{saar1998} derives that old, slow-rotating stars 
      ($v\sin{i}\leq$2 kms$^{-1}$, P$_{rot}\geq$15 d) generally have very low mean activity-related radial velocity noise 
      (<$\sigma_{v}^{\prime}$>$\sim$4.6 ms$^{-1}$). From spectroscopic analysis (see 
      Sect. \ref{sec:new-results-stars} and Table \ref{table:stars-newderivation}) 
      we have found for HD 233832 a projected velocity of 
      $v\sin{i}$=$0.8$ kms$^{-1}$ assuming $v_{macro}$=$1.8$ kms$^{-1}$; using as suggested in \cite{saar1997} the relation 
      $f_s\sim0.4\Delta y$ being the Str\"{o}mgren index $\Delta y=0.518\pm0.008$ mag for HD 233832 \cite[see][]{mints2017} we 
      obtain $f_s\sim0.21$\%. We therefore find a spot-related perturbation $A_s=1.26$ ms$^{-1}$, a bisector velocity span 
      variation $A_c=0.013$ ms$^{-1}$ and a weighted velocity dispersion $\sigma_{v}^{\prime}=3.39$ ms$^{-1}$, all well 
      below our observed $\sim80$ ms$^{-1}$ variation.
      \par To also account for radial velocity variations induced by long-term magnetic activity cycles 
      \citep[see][]{santos2010a,lovis2011,dumusque2011} that may mimic the presence of a long-period giant planet we instead use the 
      empirical models found in \cite{lovis2011} providing a relationship between effective temperature, metallicity and the measured 
      $\log R^{\prime}_{\rm HK}$ semiamplitude. The latter being 0.055 for our observations of HD 233832 (see bottom right panel of 
      Fig. \ref{fig:timeseries-plot}) we then find an estimate of the radial velocity semiamplitude induced by the stellar magnetic 
      activity cycle of $A_{RV}\sim0.30$ ms$^{-1}$, also much lower than the observed radial velocity variation of HD 233832.
      
      \paragraph{}
      While caution is certainly needed in announcing the existence of exoplanets, we propose that the signals found in the radial 
      velocity timeseries of stars HD 220197 and HD 233832 are best explained by the existence of the giant planets described in the 
      previous paragraphs.
      \par Due to its low semiamplitude of $\sim4$ ms$^{-1}$, the signal found in the data collected for star 
      HD 220197 would clearly benefit from further observation and analysis; however we find this signal to be uncorrelated with any 
      significant stellar activity indexes and that a planetary origin remains the most likely explanation given the analysis at 
      hand.
      \par Considering instead HD 233832, while we cannot completely rule out the presence of a magnetic activity cycle at about 
      either 2000 or 4000 days, we argue that a planetary origin for the radial velocity signal observed for HD 233832 seems to be 
      more likely than one exclusively induced by stellar activity since the amplitude of activity-induced radial velocity 
      variations are proven to be much lower than the one we observe. The possibility of a long-term activity cycle for the star 
      HD 233832 having a similar period to the orbital period of planet b still remains, a situation similar to the $\sim$11 years 
      period of both Jupiter and the magnetic cycle of the Sun in the Solar System that deserves future investigation.
  
  \section{Planetary frequency} \label{sec:frequency}
  
      \begin{figure}
	\includegraphics[width=\linewidth]{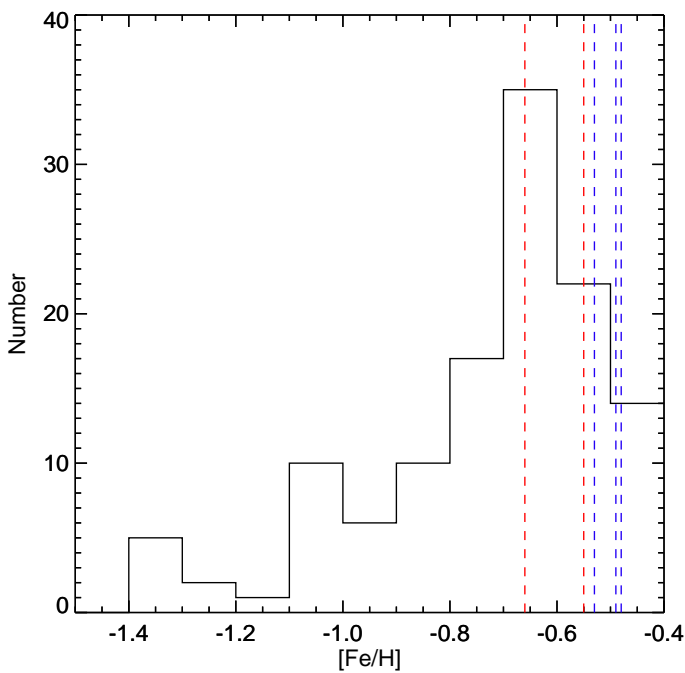}
	  \caption{Metallicity distribution for the 130 metal-poor stars, distributed in the two hemispheres. 
		    The metallicities of stars HD 220197 and HD 233832 are indicated by the vertical dashed red lines, while the 
		    metallicities of southern host stars HD 171028, HD 181720 and 190984 are shown as blue dashed lines 
		    \citep[see][]{santos2011}.}
	  \label{fig:feh-hist-tot}     
      \end{figure}
        
    Joining the two detections over our 42-stars northern sample with the three detections and one candidate in the 88-stars southern 
    sample analysed in \cite{santos2007,santos2010b,santos2011}, we obtain a total of five detected giant planets (HD 171028 b, 
    HD 181720 b, HD 190984 b, HD 220197 b and HD 233832 b) and one candidate (around HD 107094) over 130 metal-poor stars. 
    In Fig. \ref{fig:feh-hist-tot} the metallicity distribution of the overall sample is shown 
    (values from \cite{santos2011} for the southern sample and from \cite{sozzetti2009} for the northern sample), and it can be 
    noted that all 
    detected planets are found in stars in the metal-rich end of the sample; the same applies to the candidate around HD 107094 having 
    a stellar [Fe/H]=$-$0.51 dex. We also note that the planets HD 220197 b and HD 233832 b presented in this work orbit 
    the lowest 
    metallicity stars around which detected planets are found in the combined HARPS+HARPS-N sample.
    \par While an accurate 
    estimate of the frequency $f_p$ of giant planets around metal-poor stars would require an assessment of the survey detection 
    limits and will be the subject of a future paper, 
    we can preliminarily note that the 5 detected planets all have high radial velocity semiamplitude, ranging from 
    the K$\sim$4 ms$^{-1}$ signal of HD 220197 b to the K$\sim$60 ms$^{-1}$ signal of HD 171028 b; the same also applies to the proposed 
    K$\sim$88 ms$^{-1}$ planetary signal for the star HD 107094 \cite[see][]{santos2011}. The orbital periods are also long, 
    ranging from the P$\sim$550 d of HD 171028 b to the P$\sim$4885 d of HD 190984 b. We can therefore assume that such 
    combinations of high-value semiamplitude and periods would be detectable 100\% of the times with the high-precision of HARPS and 
    HARPS-N, with the possible exception of HD 220197 b having the lowest K of all detected signals.
    \par We can therefore provide a preliminar estimate of the occurrence frequency of giant planets in our metal-poor sample 
    $f_p$ using the binomial distribution:
      \begin{equation}	\label{eq:binomial}
	p(m;N,f_p)=\frac{N!}{m!(N-m)!}\ f_p^m (1-f_p)^{N-m}
      \end{equation}
    being $N=130$ our whole two-hemisphere sample assuming a detection completeness for the semiamplitudes and periods of our planetary 
    signals and $m$ the number of detections in the sample. By considering only the 5 detected giant planets we thus obtain a frequency of 
    $f_p=3.84_{-1.06}^{+2.45}\%$, while by including also the candidate signal found for star HD 107094 we have $m=6$ and a frequency 
    $f_p=4.61_{-1.21}^{+2.58}\%$, the uncertainty in both values being the 1-$\sigma$ error bar. We note that both frequency values 
    are compatible and slightly better constrained than the $3.4_{-1.0}^{+3.2}\%$ value obtained by \cite{santos2011} for their 
    three detections in the 88-stars southern sample.
    \par We stress that the detected planets all lie in the metal-rich end of the two-hemisphere sample, ranging from 
    [Fe/H]$\sim-$0.7 dex to $-$0.4 dex. Considering only the 67 stars in this metallicity range, we obtain a frequency of 
    $f_p=7.46_{-2.07}^{+4.53}\%$, comparable with the $\sim5\%$ estimate for [Fe/H]$\sim0$ dex obtained by \cite{mortier2012}, 
    \cite{mortier2013} and \cite{gonzalez2014}. 
    If we instead consider the 63 stars in the metal-poor end of the sample ([Fe/H]$<-0.7$ dex) in which 
    no giant planets were found in the two-hemisphere sample, we obtain an upper limit on frequency of $f_P<1.76\%$, significantly 
    lower than the other values obtained by our analysis again showing $f_p$ to be a rising function of [Fe/H] even at this low 
    metallicity values.
  
  \section{Summary and discussion} \label{sec:conclusions}
  
    In this work we report the detection of the two long-period giant planets HD 220197 b and HD 233832 b as a result of the intense 
    observation of 42 metal-poor stars with HARPS-N conducted as a complement to the study of the southern metal-poor sample 
    previously detailed in \cite{santos2007,santos2010b,santos2011} and to continue the analysis of the correlation between stellar 
    metallicity and giant planet frequency \citep{gonzalez1997,sozzetti2004,sozzetti2009,mortier2012,santos2011}. This correlation 
    is usually proposed as strong evidence in favour of core-accretion formation for giant planets over disk instability processes.
    \par We have characterized in Sect. \ref{sec:new-results-stars} the host stars as $\alpha$-enriched stars likely members of the 
    thick-disk stellar population; this agrees with previous studies \citep{adibekyan2012a,adibekyan2012c} finding that metal-poor 
    host stars tend to have a significant overabundance of $\alpha$-elements and be part of the thick-disk population compared to 
    non-hosting stars; in addition \cite{maldonado2018} also notes that stars hosting cool Jupiters like HD 220197 b and HD 233832 b 
    also have higher $\alpha$ abundances than stars hosting hot Jupiters, suggesting that the low [Fe/H] content in such 
    protoplanetary disks may be compensated by this overabundance of $\alpha$-elements allowing the formation of planetary cores.    
    \par We find HD 220197 b to be characterized as a M$\sin{i}=0.20_{-0.04}^{+0.07}$ M$_J$ orbiting its host star 
    with a period of $1728_{-80}^{+162}$ days; althought it could be noted that this low-amplitude signal 
    (K=$3.78_{-0.72}^{+1.78}$ ms$^{-1}$) should be treated with caution, we find that given the results of the current analysis the 
    planetary nature of the signal remains the most likely explanation. We note that this planet is the least massive 
    long-period ($P>1$ yr) giant planet found around such a metal-poor star, an interesting counter-example to the tendency of metal-poor stars 
    to host more massive planets \cite[e.g.][]{santos2017}.
    \par On the other hand, the orbital characteristics of 
    HD 233832 are more ambiguous, finding two solutions of equal statistical weight at 
    P=$2058_{-40}^{+47}$ d and P=$4047_{-117}^{+91}$, returning possible minimum masses of $1.78_{-0.06}^{+0.08}$ M$_J$ 
    and $2.72_{-0.23}^{+0.23}$ M$_J$, clearly needing more radial velocity data providing a better sampling of its Doppler variation to 
    discriminate between the two competing solutions. Since the analysis of stellar activity is always an important part in 
    searching for planetary signals, extra special care was taken in handling the signal found for HD 233832 due to the presence of 
    a long-period activity signal in the periodogram of activity index $\log R^{\prime}_{\rm HK}$ near our proposed orbital 
    solutions. Following 
    the analysis of activity-induced radial velocity signals of \cite{saar1997,saar1998,saar2000,santos2010a,lovis2011,dumusque2011} 
    we argue however that such activity 
    signals would be at most of the order of 5 ms$^{-1}$, well below the $\sim$80 ms$^{-1}$ variation observed for HD 233832, 
    suggesting a planetary origin for the radial velocity signal. The same applies for the influence of its stellar companion at 
    angular separation of 4.8 arcseconds recently detected by Gaia. Not being however able to completely rule out the presence of a 
    low-amplitude activity cycle having period similar to our solutions for HD 233832 b, similarly to what is found in the Solar System 
    for the orbital period of Jupiter and the magnetic activity cycle of the Sun, more analysis will be needed to fully 
    characterize this interesting planetary system.
    \par Joining our detections with the three giant planets (HD 171028 b, HD 181720 b and HD 190984 b) detected in the HARPS 
    metal-poor sample \citep[see][]{santos2011} we obtain a total of five detections over 130 metal-poor stars. In a preliminary 
    statistical analysis, assuming survey 
    completeness for the high-value radial velocity semiamplitudes (4 to 60 ms$^{-1}$) and periods (550 to 4885 days) of the detected 
    planets, we find a frequency of giant planets around metal-poor stars of $3.84_{-1.06}^{+2.45}$\%; this values rises to 
    $4.61_{-1.21}^{+2.58}$\% when including also the candidate planetary signal found in the southern sample around HD 107094 
    \citep[see][]{santos2011}. If we instead consider only the 67 stars in the metallicity range ($-$0.7 <[Fe/H]< $-$0.4\,dex) in 
    which the detected 
    planets are found, the frequency rises to $f_p=7.46_{-2.07}^{+4.53}\%$, a value similar to literature estimates on giant 
    planets frequency around solar-metallicity stars \citep[see][]{mortier2012,mortier2013,gonzalez2014}. We stress that, 
    similarly to the case of the \cite{santos2011} analysis, our frequency results are conservative as we do not account for 
    completeness in the survey, which will be the subject of a future paper.
    \par Interestingly, the host stars lie in the metal-rich end of the overall stellar sample, reinforcing previous results 
    suggesting that frequency of giant planets continues to be a rising function of stellar metallicity even for metal-poor stars and 
    favouring core-accretion processes for the formation of giant planets \citep{ida2004,mordasini2009a}; this is also reinforced by 
    the $f_p<1.76\%$ obtained from the null detections below [Fe/H]$<-0.7$ dex. We however note that our calculation of $f_p$ should 
    be seen as a preliminar update of giant planets frequency around metal-poor stars, and that a more rigorous assessment of its 
    value will be the subject of a future paper.
    \par While the correlation between stellar metallicity and occurrence of giant planets continues to be confirmed by observations, 
    more analysis is clearly needed to provide a more solid observational basis on which to shed light on planet formation mechanisms
    and their relation to host star characteristics.
    
    \begin{acknowledgements}
      We thank the anonymous referee for the useful comments.
      The GAPS project acknowledges the support by INAF/Frontiera through the "Progetti Premiali" funding scheme of the Italian 
      Ministry of Education, University, and Research. 
      DB acknowledges financial support from INAF and Agenzia Spaziale Italiana (ASI grant n. 014-025-R.1.2015) for the 2016 PhD 
      fellowship programme of INAF.
      NCS and JF acknowledge the support by the Fundação para a Ciência e Tecnologia (FCT) through national funds and by FEDER through 
      COMPETE2020 by grants UID/FIS/04434/2013 \& POCI-01-0145-FEDER-007672. This work was also funded by FEDER - Fundo Europeu de 
      Desenvolvimento Regional funds through the COMPETE 2020 - Programa Operacional Competitividade e Internacionalização (POCI), 
      and by Portuguese funds through FCT in the framework of the project POCI-01-0145-FEDER-028953 and POCI-01-0145-FEDER-032113.
      L.M. acknowledges support from the Italian Minister of Instruction, University and Research (MIUR) through FFABR 2017 fund. 
      L.M. also acknowledges support from the University of Rome Tor Vergata through "Mission: Sustainability 2016" fund. 
      G.S. acknowledges financial support from “Accordo ASI–INAF” No. 2013-016-R.0 July 9, 2013 and July 9, 2015.
    \end{acknowledgements}
  
  \bibliographystyle{aa}
  \bibliography{ref}

\begin{thebibliography}{99}
\expandafter\ifx\csname natexlab\endcsname\relax\def\natexlab#1{#1}\fi

\bibitem[{{Adibekyan} {et~al.}(2012{\natexlab{a}}){Adibekyan}, {Delgado Mena},
  {Sousa}, {Santos}, {Israelian}, {Gonz{\'a}lez Hern{\'a}ndez}, {Mayor}, \&
  {Hakobyan}}]{adibekyan2012c}
{Adibekyan}, V.~Z., {Delgado Mena}, E., {Sousa}, S.~G., {et~al.}
  2012{\natexlab{a}}, \aap, 547, A36

\bibitem[{{Adibekyan} {et~al.}(2012{\natexlab{b}}){Adibekyan}, {Santos},
  {Sousa}, {Israelian}, {Delgado Mena}, {Gonz{\'a}lez Hern{\'a}ndez}, {Mayor},
  {Lovis}, \& {Udry}}]{adibekyan2012a}
{Adibekyan}, V.~Z., {Santos}, N.~C., {Sousa}, S.~G., {et~al.}
  2012{\natexlab{b}}, \aap, 543, A89

\bibitem[{{Adibekyan} {et~al.}(2012{\natexlab{c}}){Adibekyan}, {Sousa},
  {Santos}, {Delgado Mena}, {Gonz{\'a}lez Hern{\'a}ndez}, {Israelian}, {Mayor},
  \& {Khachatryan}}]{adibekyan2012b}
{Adibekyan}, V.~Z., {Sousa}, S.~G., {Santos}, N.~C., {et~al.}
  2012{\natexlab{c}}, \aap, 545, A32

\bibitem[{{Benatti} {et~al.}(2017){Benatti}, {Desidera}, {Damasso},
  {Malavolta}, {Lanza}, {Biazzo}, {Bonomo}, {Claudi}, {Marzari}, {Poretti},
  {Gratton}, {Micela}, {Pagano}, {Piotto}, {Sozzetti}, {Boccato}, {Cosentino},
  {Covino}, {Maggio}, {Molinari}, {Smareglia}, {Affer}, {Andreuzzi},
  {Bignamini}, {Borsa}, {di Fabrizio}, {Esposito}, {Martinez Fiorenzano},
  {Messina}, {Giacobbe}, {Harutyunyan}, {Knapic}, {Maldonado}, {Masiero},
  {Nascimbeni}, {Pedani}, {Rainer}, {Scandariato}, \& {Silvotti}}]{benatti2017}
{Benatti}, S., {Desidera}, S., {Damasso}, M., {et~al.} 2017, \aap, 599, A90

\bibitem[{{Bensby} {et~al.}(2003){Bensby}, {Feltzing}, \&
  {Lundstr{\"o}m}}]{bensby2003}
{Bensby}, T., {Feltzing}, S., \& {Lundstr{\"o}m}, I. 2003, \aap, 410, 527

\bibitem[{{Bensby} {et~al.}(2005){Bensby}, {Feltzing}, {Lundstr{\"o}m}, \&
  {Ilyin}}]{bensby2005}
{Bensby}, T., {Feltzing}, S., {Lundstr{\"o}m}, I., \& {Ilyin}, I. 2005, \aap,
  433, 185

\bibitem[{{Biazzo} {et~al.}(2015){Biazzo}, {Gratton}, {Desidera}, {Lucatello},
  {Sozzetti}, {Bonomo}, {Damasso}, {Gandolfi}, {Affer}, {Boccato}, {Borsa},
  {Claudi}, {Cosentino}, {Covino}, {Knapic}, {Lanza}, {Maldonado}, {Marzari},
  {Micela}, {Molaro}, {Pagano}, {Pedani}, {Pillitteri}, {Piotto}, {Poretti},
  {Rainer}, {Santos}, {Scandariato}, \& {Zanmar Sanchez}}]{biazzo2015}
{Biazzo}, K., {Gratton}, R., {Desidera}, S., {et~al.} 2015, \aap, 583

\bibitem[{{Biazzo} {et~al.}(2007){Biazzo}, {Pasquini}, {Girardi}, {Frasca}, {da
  Silva}, {Setiawan}, {Marilli}, {Hatzes}, \& {Catalano}}]{biazzo2007}
{Biazzo}, K., {Pasquini}, L., {Girardi}, L., {et~al.} 2007, \aap, 475, 981

\bibitem[{{Bonfils} {et~al.}(2013){Bonfils}, {Delfosse}, {Udry}, {Forveille},
  {Mayor}, {Perrier}, {Bouchy}, {Gillon}, {Lovis}, {Pepe}, {Queloz}, {Santos},
  {S{\'e}gransan}, \& {Bertaux}}]{bonfils2013}
{Bonfils}, X., {Delfosse}, X., {Udry}, S., {et~al.} 2013, \aap, 549, A109

\bibitem[{{Boss}(1997)}]{boss1997}
{Boss}, A.~P. 1997, Science, 276, 1836

\bibitem[{{Boss}(2002)}]{boss2002}
{Boss}, A.~P. 2002, \apj, 576, 462

\bibitem[{{Boss}(2006)}]{boss2006}
{Boss}, A.~P. 2006, \apjl, 644, L79

\bibitem[{{Brewer} {et~al.}(2016){Brewer}, {Fischer}, {Valenti}, \&
  {Piskunov}}]{brewer2016}
{Brewer}, J.~M., {Fischer}, D.~A., {Valenti}, J.~A., \& {Piskunov}, N. 2016,
  The Astrophysical Journal Supplement Series, 225, 32

\bibitem[{{Brown} {et~al.}(2018){Brown}, {Vallenari}, {Prusti}, {de Bruijne},
  {Babusiaux}, {Bailer-Jones}, {Biermann}, {Evans}, {Eyer}, {Jansen}, {Jordi},
  {Klioner}, {Lammers}, {Lindegren}, {Luri}, {Mignard}, {Panem}, {Pourbaix},
  {Randich}, {Sartoretti}, {Siddiqui}, {Soubiran}, {van Leeuwen}, {Walton},
  {Arenou}, {Bastian}, {Cropper}, {Drimmel}, {Katz}, {Lattanzi}, {Bakker},
  {Cacciari}, {Casta{\~n}eda}, {Chaoul}, {Cheek}, {De Angeli}, {Fabricius},
  {Guerra}, {Holl}, {Masana}, {Messineo}, {Mowlavi}, {Nienartowicz}, {Panuzzo},
  {Portell}, {Riello}, {Seabroke}, {Tanga}, {Th{\'e}venin}, {Gracia-Abril},
  {Comoretto}, {Garcia-Reinaldos}, {Teyssier}, {Altmann}, {Andrae}, {Audard},
  {Bellas-Velidis}, {Benson}, {Berthier}, {Blomme}, {Burgess}, {Busso},
  {Carry}, {Cellino}, {Clementini}, {Clotet}, {Creevey}, {Davidson}, {De
  Ridder}, {Delchambre}, {Dell'Oro}, {Ducourant},
  {Fern{\'a}ndez-Hern{\'a}ndez}, {Fouesneau}, {Fr{\'e}mat}, {Galluccio},
  {Garc{\'\i}a-Torres}, {Gonz{\'a}lez-N{\'u}{\~n}ez}, {Gonz{\'a}lez- Vidal},
  {Gosset}, {Guy}, {Halbwachs}, {Hambly}, {Harrison}, {Hern{\'a}ndez},
  {Hestroffer}, {Hodgkin}, {Hutton}, {Jasniewicz}, {Jean-Antoine- Piccolo},
  {Jordan}, {Korn}, {Krone- Martins}, {Lanzafame}, {Lebzelter}, {L{\"o}ffler},
  {Manteiga}, {Marrese}, {Mart{\'\i}n-Fleitas}, {Moitinho}, {Mora}, {Muinonen},
  {Osinde}, {Pancino}, {Pauwels}, {Petit}, {Recio-Blanco}, {Richards},
  {Rimoldini}, {Robin}, {Sarro}, {Siopis}, {Smith}, {Sozzetti}, {S{\"u}veges},
  {Torra}, {van Reeven}, {Abbas}, {Abreu Aramburu}, {Accart}, {Aerts},
  {Altavilla}, {{\'A}lvarez}, {Alvarez}, {Alves}, {Anderson}, {Andrei},
  {Anglada Varela}, {Antiche}, {Antoja}, {Arcay}, {Astraatmadja}, {Bach},
  {Baker}, {Balaguer-N{\'u}{\~n}ez}, {Balm}, {Barache}, {Barata}, {Barbato},
  {Barblan}, {Barklem}, {Barrado}, {Barros}, {Barstow}, {Bartholom{\'e}
  Mu{\~n}oz}, {Bassilana}, {Becciani}, {Bellazzini}, {Berihuete}, {Bertone},
  {Bianchi}, {Bienaym{\'e}}, {Blanco-Cuaresma}, {Boch}, {Boeche}, {Bombrun},
  {Borrachero}, {Bossini}, {Bouquillon}, {Bourda}, {Bragaglia}, {Bramante},
  {Breddels}, {Bressan}, {Brouillet}, {Br{\"u}semeister}, {Brugaletta},
  {Bucciarelli}, {Burlacu}, {Busonero}, {Butkevich}, {Buzzi}, {Caffau},
  {Cancelliere}, {Cannizzaro}, {Cantat-Gaudin}, {Carballo}, {Carlucci},
  {Carrasco}, {Casamiquela}, {Castellani}, {Castro-Ginard}, {Charlot},
  {Chemin}, {Chiavassa}, {Cocozza}, {Costigan}, {Cowell}, {Crifo}, {Crosta},
  {Crowley}, {Cuypers}, {Dafonte}, {Damerdji}, {Dapergolas}, {David}, {David},
  {de Laverny}, {De Luise}, {De March}, {de Martino}, {de Souza}, {de Torres},
  {Debosscher}, {del Pozo}, {Delbo}, {Delgado}, {Delgado}, {Di Matteo},
  {Diakite}, {Diener}, {Distefano}, {Dolding}, {Drazinos}, {Dur{\'a}n},
  {Edvardsson}, {Enke}, {Eriksson}, {Esquej}, {Eynard Bontemps}, {Fabre},
  {Fabrizio}, {Faigler}, {Falc{\~a}o}, {Farr{\`a}s Casas}, {Federici},
  {Fedorets}, {Fernique}, {Figueras}, {Filippi}, {Findeisen}, {Fonti},
  {Fraile}, {Fraser}, {Fr{\'e}zouls}, {Gai}, {Galleti}, {Garabato},
  {Garc{\'\i}a-Sedano}, {Garofalo}, {Garralda}, {Gavel}, {Gavras}, {Gerssen},
  {Geyer}, {Giacobbe}, {Gilmore}, {Girona}, {Giuffrida}, {Glass}, {Gomes},
  {Granvik}, {Gueguen}, {Guerrier}, {Guiraud}, {Guti{\'e}rrez-S{\'a}nchez},
  {Haigron}, {Hatzidimitriou}, {Hauser}, {Haywood}, {Heiter}, {Helmi}, {Heu},
  {Hilger}, {Hobbs}, {Hofmann}, {Holland}, {Huckle}, {Hypki}, {Icardi},
  {Jan{\ss}en}, {Jevardat de Fombelle}, {Jonker}, {Juh{\'a}sz}, {Julbe},
  {Karampelas}, {Kewley}, {Klar}, {Kochoska}, {Kohley}, {Kolenberg},
  {Kontizas}, {Kontizas}, {Koposov}, {Kordopatis}, {Kostrzewa-Rutkowska},
  {Koubsky}, {Lambert}, {Lanza}, {Lasne}, {Lavigne}, {Le Fustec}, {Le
  Poncin-Lafitte}, {Lebreton}, {Leccia}, {Leclerc}, {Lecoeur-Taibi},
  {Lenhardt}, {Leroux}, {Liao}, {Licata}, {Lindstr{\o}m}, {Lister}, {Livanou},
  {Lobel}, {L{\'o}pez}, {Managau}, {Mann}, {Mantelet}, {Marchal}, {Marchant},
  {Marconi}, {Marinoni}, {Marschalk{\'o}}, {Marshall}, {Martino}, {Marton},
  {Mary}, {Massari}, {Matijevi{\v{c}}}, {Mazeh}, {McMillan}, {Messina},
  {Michalik}, {Millar}, {Molina}, {Molinaro}, {Moln{\'a}r}, {Montegriffo},
  {Mor}, {Morbidelli}, {Morel}, {Morris}, {Mulone}, {Muraveva}, {Musella},
  {Nelemans}, {Nicastro}, {Noval}, {O'Mullane}, {Ord{\'e}novic},
  {Ord{\'o}{\~n}ez-Blanco}, {Osborne}, {Pagani}, {Pagano}, {Pailler},
  {Palacin}, {Palaversa}, {Panahi}, {Pawlak}, {Piersimoni}, {Pineau}, {Plachy},
  {Plum}, {Poggio}, {Poujoulet}, {Pr{\v{s}}a}, {Pulone}, {Racero}, {Ragaini},
  {Rambaux}, {Ramos-Lerate}, {Regibo}, {Reyl{\'e}}, {Riclet}, {Ripepi}, {Riva},
  {Rivard}, {Rixon}, {Roegiers}, {Roelens}, {Romero-G{\'o}mez}, {Rowell},
  {Royer}, {Ruiz-Dern}, {Sadowski}, {Sagrist{\`a} Sell{\'e}s}, {Sahlmann},
  {Salgado}, {Salguero}, {Sanna}, {Santana- Ros}, {Sarasso}, {Savietto},
  {Schultheis}, {Sciacca}, {Segol}, {Segovia}, {S{\'e}gransan}, {Shih},
  {Siltala}, {Silva}, {Smart}, {Smith}, {Solano}, {Solitro}, {Sordo}, {Soria
  Nieto}, {Souchay}, {Spagna}, {Spoto}, {Stampa}, {Steele},
  {Steidelm{\"u}ller}, {Stephenson}, {Stoev}, {Suess}, {Surdej}, {Szabados},
  {Szegedi-Elek}, {Tapiador}, {Taris}, {Tauran}, {Taylor}, {Teixeira},
  {Terrett}, {Teyssandier}, {Thuillot}, {Titarenko}, {Torra Clotet}, {Turon},
  {Ulla}, {Utrilla}, {Uzzi}, {Vaillant}, {Valentini}, {Valette}, {van Elteren},
  {Van Hemelryck}, {van Leeuwen}, {Vaschetto}, {Vecchiato}, {Veljanoski},
  {Viala}, {Vicente}, {Vogt}, {von Essen}, {Voss}, {Votruba}, {Voutsinas},
  {Walmsley}, {Weiler}, {Wertz}, {Wevers}, {Wyrzykowski}, {Yoldas},
  {{\v{Z}}erjal}, {Ziaeepour}, {Zorec}, {Zschocke}, {Zucker}, {Zurbach}, \&
  {Zwitter}}]{gaia2018}
{Brown}, A.~G.~A., {Vallenari}, A., {Prusti}, T., {et~al.} 2018, \aap, 616, A1

\bibitem[{{Buchhave} {et~al.}(2018){Buchhave}, {Bitsch}, {Johansen}, {Latham},
  {Bizzarro}, {Bieryla}, \& {Kipping}}]{buchhave2018}
{Buchhave}, L.~A., {Bitsch}, B., {Johansen}, A., {et~al.} 2018, \apj, 856, 37

\bibitem[{{Buchhave} {et~al.}(2012){Buchhave}, {Latham}, {Johansen},
  {Bizzarro}, {Torres}, {Rowe}, {Batalha}, {Borucki}, {Brugamyer}, {Caldwell},
  {Bryson}, {Ciardi}, {Cochran}, {Endl}, {Esquerdo}, {Ford}, {Geary},
  {Gilliland}, {Hansen}, {Isaacson}, {Laird}, {Lucas}, {Marcy}, {Morse},
  {Robertson}, {Shporer}, {Stefanik}, {Still}, \& {Quinn}}]{buchhave2012}
{Buchhave}, L.~A., {Latham}, D.~W., {Johansen}, A., {et~al.} 2012, \nat, 486,
  375

\bibitem[{{Butler} {et~al.}(2004){Butler}, {Vogt}, {Marcy}, {Fischer},
  {Wright}, {Henry}, {Laughlin}, \& {Lissauer}}]{butler2004}
{Butler}, R.~P., {Vogt}, S.~S., {Marcy}, G.~W., {et~al.} 2004, \apj, 617, 580

\bibitem[{{Carney} {et~al.}(1996){Carney}, {Laird}, {Latham}, \&
  {Aguilar}}]{carney1996}
{Carney}, B.~W., {Laird}, J.~B., {Latham}, D.~W., \& {Aguilar}, L.~A. 1996,
  \aj, 112, 668

\bibitem[{{Carolo} {et~al.}(2014){Carolo}, {Desidera}, {Gratton}, {Martinez
  Fiorenzano}, {Marzari}, {Endl}, {Mesa}, {Barbieri}, {Cecconi}, {Claudi},
  {Cosentino}, \& {Scuderi}}]{carolo2014}
{Carolo}, E., {Desidera}, S., {Gratton}, R., {et~al.} 2014, \aap, 567, A48

\bibitem[{{Cosentino} {et~al.}(2012){Cosentino}, {Lovis}, {Pepe}, {Collier
  Cameron}, {Latham}, {Molinari}, {Udry}, {Bezawada}, {Black}, {Born},
  {Buchschacher}, {Charbonneau}, {Figueira}, {Fleury}, {Galli}, {Gallie},
  {Gao}, {Ghedina}, {Gonzalez}, {Gonzalez}, {Guerra}, {Henry}, {Horne},
  {Hughes}, {Kelly}, {Lodi}, {Lunney}, {Maire}, {Mayor}, {Micela}, {Ordway},
  {Peacock}, {Phillips}, {Piotto}, {Pollacco}, {Queloz}, {Rice}, {Riverol},
  {Riverol}, {San Juan}, {Sasselov}, {Segransan}, {Sozzetti}, {Sosnowska},
  {Stobie}, {Szentgyorgyi}, {Vick}, \& {Weber}}]{cosentino2012}
{Cosentino}, R., {Lovis}, C., {Pepe}, F., {et~al.} 2012, in Ground-based and
  Airborne Instrumentation for Astronomy IV, Vol. 8446, 84461V

\bibitem[{{Courcol} {et~al.}(2016){Courcol}, {Bouchy}, \&
  {Deleuil}}]{courcol2016}
{Courcol}, B., {Bouchy}, F., \& {Deleuil}, M. 2016, \mnras, 461, 1841

\bibitem[{{Covino} {et~al.}(2013){Covino}, {Esposito}, {Barbieri}, {Mancini},
  {Nascimbeni}, {Claudi}, {Desidera}, {Gratton}, {Lanza}, {Sozzetti}, {Biazzo},
  {Affer}, {Gandolfi}, {Munari}, {Pagano}, {Bonomo}, {Collier Cameron},
  {H{\'e}brard}, {Maggio}, {Messina}, {Micela}, {Molinari}, {Pepe}, {Piotto},
  {Ribas}, {Santos}, {Southworth}, {Shkolnik}, {Triaud}, {Bedin}, {Benatti},
  {Boccato}, {Bonavita}, {Borsa}, {Borsato}, {Brown}, {Carolo}, {Ciceri},
  {Cosentino}, {Damasso}, {Faedi}, {Mart{\'{\i}}nez Fiorenzano}, {Latham},
  {Lovis}, {Mordasini}, {Nikolov}, {Poretti}, {Rainer}, {Rebolo L{\'o}pez},
  {Scandariato}, {Silvotti}, {Smareglia}, {Alcal{\'a}}, {Cunial}, {Di
  Fabrizio}, {Di Mauro}, {Giacobbe}, {Granata}, {Harutyunyan}, {Knapic},
  {Lattanzi}, {Leto}, {Lodato}, {Malavolta}, {Marzari}, {Molinaro},
  {Nardiello}, {Pedani}, {Prisinzano}, \& {Turrini}}]{covino2013}
{Covino}, E., {Esposito}, M., {Barbieri}, M., {et~al.} 2013, \aap, 554, A28

\bibitem[{{Cutri} {et~al.}(2003){Cutri}, {Skrutskie}, {van Dyk}, {Beichman},
  {Carpenter}, {Chester}, {Cambresy}, {Evans}, {Fowler}, {Gizis}, {Howard},
  {Huchra}, {Jarrett}, {Kopan}, {Kirkpatrick}, {Light}, {Marsh}, {McCallon},
  {Schneider}, {Stiening}, {Sykes}, {Weinberg}, {Wheaton}, {Wheelock}, \&
  {Zacarias}}]{cutri2003}
{Cutri}, R.~M., {Skrutskie}, M.~F., {van Dyk}, S., {et~al.} 2003, VizieR Online
  Data Catalog, II/246

\bibitem[{{da Silva} {et~al.}(2006){da Silva}, {Girardi}, {Pasquini},
  {Setiawan}, {von der L{\"u}he}, {de Medeiros}, {Hatzes}, {D{\"o}llinger}, \&
  {Weiss}}]{dasilva2006}
{da Silva}, L., {Girardi}, L., {Pasquini}, L., {et~al.} 2006, \aap, 458, 609

\bibitem[{{Damasso} {et~al.}(2015){Damasso}, {Biazzo}, {Bonomo}, {Desidera},
  {Lanza}, {Nascimbeni}, {Esposito}, {Scandariato}, {Sozzetti}, {Cosentino},
  {Gratton}, {Malavolta}, {Rainer}, {Gandolfi}, {Poretti}, {Zanmar Sanchez},
  {Ribas}, {Santos}, {Affer}, {Andreuzzi}, {Barbieri}, {Bedin}, {Benatti},
  {Bernagozzi}, {Bertolini}, {Bonavita}, {Borsa}, {Borsato}, {Boschin},
  {Calcidese}, {Carbognani}, {Cenadelli}, {Christille}, {Claudi}, {Covino},
  {Cunial}, {Giacobbe}, {Granata}, {Harutyunyan}, {Lattanzi}, {Leto},
  {Libralato}, {Lodato}, {Lorenzi}, {Mancini}, {Martinez Fiorenzano},
  {Marzari}, {Masiero}, {Micela}, {Molinari}, {Molinaro}, {Munari}, {Murabito},
  {Pagano}, {Pedani}, {Piotto}, {Rosenberg}, {Silvotti}, \&
  {Southworth}}]{damasso2015}
{Damasso}, M., {Biazzo}, K., {Bonomo}, A.~S., {et~al.} 2015, \aap, 575

\bibitem[{{Desidera} {et~al.}(2014){Desidera}, {Bonomo}, {Claudi}, {Damasso},
  {Biazzo}, {Sozzetti}, {Marzari}, {Benatti}, {Gandolfi}, {Gratton}, {Lanza},
  {Nascimbeni}, {Andreuzzi}, {Affer}, {Barbieri}, {Bedin}, {Bignamini},
  {Bonavita}, {Borsa}, {Calcidese}, {Christille}, {Cosentino}, {Covino},
  {Esposito}, {Giacobbe}, {Harutyunyan}, {Latham}, {Lattanzi}, {Leto},
  {Lodato}, {Lovis}, {Maggio}, {Malavolta}, {Mancini}, {Martinez Fiorenzano},
  {Micela}, {Molinari}, {Mordasini}, {Munari}, {Pagano}, {Pedani}, {Pepe},
  {Piotto}, {Poretti}, {Rainer}, {Ribas}, {Santos}, {Scandariato}, {Silvotti},
  {Southworth}, \& {Zanmar Sanchez}}]{desidera2004}
{Desidera}, S., {Bonomo}, A.~S., {Claudi}, R.~U., {et~al.} 2014, \aap, 567, L6

\bibitem[{{Desidera} {et~al.}(2013){Desidera}, {Sozzetti}, {Bonomo}, {Gratton},
  {Poretti}, {Claudi}, {Latham}, {Affer}, {Cosentino}, {Damasso}, {Esposito},
  {Giacobbe}, {Malavolta}, {Nascimbeni}, {Piotto}, {Rainer}, {Scardia},
  {Schmid}, {Lanza}, {Micela}, {Pagano}, {Bedin}, {Biazzo}, {Borsa}, {Carolo},
  {Covino}, {Faedi}, {H{\'e}brard}, {Lovis}, {Maggio}, {Mancini}, {Marzari},
  {Messina}, {Molinari}, {Munari}, {Pepe}, {Santos}, {Scandariato}, {Shkolnik},
  \& {Southworth}}]{desidera2013}
{Desidera}, S., {Sozzetti}, A., {Bonomo}, A.~S., {et~al.} 2013, \aap, 554, A29

\bibitem[{{Dumusque} {et~al.}(2011){Dumusque}, {Lovis}, {S{\'e}gransan},
  {Mayor}, {Udry}, {Benz}, {Bouchy}, {Lo Curto}, {Mordasini}, {Pepe}, {Queloz},
  {Santos}, \& {Naef}}]{dumusque2011}
{Dumusque}, X., {Lovis}, C., {S{\'e}gransan}, D., {et~al.} 2011, \aap, 535, A55

\bibitem[{{Eastman} {et~al.}(2013){Eastman}, {Gaudi}, \& {Agol}}]{eastman2013}
{Eastman}, J., {Gaudi}, B.~S., \& {Agol}, E. 2013, \pasp, 125, 83

\bibitem[{{Endl} {et~al.}(2016){Endl}, {Brugamyer}, {Cochran}, {MacQueen},
  {Robertson}, {Meschiari}, {Ramirez}, {Shetrone}, {Gullikson}, {Johnson},
  {Wittenmyer}, {Horner}, {Ciardi}, {Horch}, {Simon}, {Howell}, {Everett},
  {Caldwell}, \& {Castanheira}}]{endl2016}
{Endl}, M., {Brugamyer}, E.~J., {Cochran}, W.~D., {et~al.} 2016, \apj, 818, 34

\bibitem[{{Figueira} {et~al.}(2013){Figueira}, {Santos}, {Pepe}, {Lovis}, \&
  {Nardetto}}]{figueira2013}
{Figueira}, P., {Santos}, N.~C., {Pepe}, F., {Lovis}, C., \& {Nardetto}, N.
  2013, \aap, 557

\bibitem[{{Figueira} {et~al.}(2015){Figueira}, {Santos}, {Pepe}, {Lovis}, \&
  {Nardetto}}]{figueira2015}
{Figueira}, P., {Santos}, N.~C., {Pepe}, F., {Lovis}, C., \& {Nardetto}, N.
  2015, \aap, 582

\bibitem[{{Fischer} \& {Valenti}(2005)}]{fischer2005}
{Fischer}, D.~A. \& {Valenti}, J. 2005, \apj, 622, 1102

\bibitem[{{Gaia Collaboration} {et~al.}(2016){Gaia Collaboration}, {Prusti},
  {de Bruijne}, {Brown}, {Vallenari}, {Babusiaux}, {Bailer-Jones}, {Bastian},
  {Biermann}, {Evans}, \& et~al.}]{gaia2016}
{Gaia Collaboration}, {Prusti}, T., {de Bruijne}, J.~H.~J., {et~al.} 2016,
  \aap, 595, A1

\bibitem[{{Gonzalez}(1997)}]{gonzalez1997}
{Gonzalez}, G. 1997, \mnras, 285, 403

\bibitem[{{Gonzalez}(2014)}]{gonzalez2014}
{Gonzalez}, G. 2014, \mnras, 443, 393

\bibitem[{{Haywood}(2008)}]{haywood2008}
{Haywood}, M. 2008, \aap, 482, 673

\bibitem[{{Haywood}(2009)}]{haywood2009}
{Haywood}, M. 2009, \apj, 698, L1

\bibitem[{{Ida} \& {Lin}(2004)}]{ida2004}
{Ida}, S. \& {Lin}, D.~N.~C. 2004, \apj, 616, 567

\bibitem[{{Johnson} {et~al.}(2010){Johnson}, {Aller}, {Howard}, \&
  {Crepp}}]{johnson2010}
{Johnson}, J.~A., {Aller}, K.~M., {Howard}, A.~W., \& {Crepp}, J.~R. 2010,
  Publications of the Astronomical Society of the Pacific, 122, 905

\bibitem[{{Johnson} {et~al.}(2007){Johnson}, {Butler}, {Marcy}, {Fischer},
  {Vogt}, {Wright}, \& {Peek}}]{johnson2007}
{Johnson}, J.~A., {Butler}, R.~P., {Marcy}, G.~W., {et~al.} 2007, \apj, 670,
  833

\bibitem[{{Johnson} \& {Li}(2012)}]{johnson2012}
{Johnson}, J.~L. \& {Li}, H. 2012, \apj, 751, 81

\bibitem[{{Johnson} {et~al.}(2016){Johnson}, {Endl}, {Cochran}, {Meschiari},
  {Robertson}, {MacQueen}, {Brugamyer}, {Caldwell}, {Hatzes}, {Ram{\'\i}rez},
  \& {Wittenmyer}}]{johnson2016}
{Johnson}, M.~C., {Endl}, M., {Cochran}, W.~D., {et~al.} 2016, \apj, 821, 74

\bibitem[{{Jones} \& {Jenkins}(2014)}]{jones2014}
{Jones}, M.~I. \& {Jenkins}, J.~S. 2014, \aap, 562, A129

\bibitem[{{Kang} {et~al.}(2011){Kang}, {Lee}, \& {Kim}}]{kang2011}
{Kang}, W., {Lee}, S.-G., \& {Kim}, K.-M. 2011, \apj, 736, 87

\bibitem[{{Kurucz}(1993)}]{kurucz1993}
{Kurucz}, R.~L. 1993, in Astronomical Society of the Pacific Conference Series,
  Vol.~44, IAU Colloq. 138: Peculiar versus Normal Phenomena in A-type and
  Related Stars, ed. M.~M. {Dworetsky}, F.~{Castelli}, \& R.~{Faraggiana}, 87

\bibitem[{{Lanza} {et~al.}(2018){Lanza}, {Malavolta}, {Benatti}, {Desidera},
  {Bignamini}, {Bonomo}, {Esposito}, {Figueira}, {Gratton}, {Scandariato},
  {Damasso}, {Sozzetti}, {Biazzo}, {Claudi}, {Cosentino}, {Covino}, {Maggio},
  {Masiero}, {Micela}, {Molinari}, {Pagano}, {Piotto}, {Poretti}, {Smareglia},
  {Affer}, {Boccato}, {Borsa}, {Boschin}, {Giacobbe}, {Knapic}, {Leto},
  {Maldonado}, {Mancini}, {Martinez Fiorenzano}, {Messina}, {Nascimbeni},
  {Pedani}, \& {Rainer}}]{lanza2018}
{Lanza}, A.~F., {Malavolta}, L., {Benatti}, S., {et~al.} 2018, A\&A, 616, A155

\bibitem[{{Lanza} {et~al.}(2016){Lanza}, {Molaro}, {Monaco}, \&
  {Haywood}}]{lanza2016}
{Lanza}, A.~F., {Molaro}, P., {Monaco}, L., \& {Haywood}, R.~D. 2016, \aap,
  587, A103

\bibitem[{{Lovis} {et~al.}(2011){Lovis}, {Dumusque}, {Santos}, {Bouchy},
  {Mayor}, {Pepe}, {Queloz}, {S{\'e}gransan}, \& {Udry}}]{lovis2011}
{Lovis}, C., {Dumusque}, X., {Santos}, N.~C., {et~al.} 2011, ArXiv e-prints
  [\eprint[arXiv]{1107.5325}]

\bibitem[{{Maldonado} {et~al.}(2015{\natexlab{a}}){Maldonado}, {Affer},
  {Micela}, {Scandariato}, {Damasso}, {Stelzer}, {Barbieri}, {Bedin}, {Biazzo},
  {Bignamini}, {Borsa}, {Claudi}, {Covino}, {Desidera}, {Esposito}, {Gratton},
  {Gonz{\'a}lez Hern{\'a}ndez}, {Lanza}, {Maggio}, {Molinari}, {Pagano},
  {Perger}, {Pillitteri}, {Piotto}, {Poretti}, {Prisinzano}, {Rebolo}, {Ribas},
  {Shkolnik}, {Southworth}, {Sozzetti}, \& {Su{\'a}rez
  Mascare{\~n}o}}]{maldonado2015a}
{Maldonado}, J., {Affer}, L., {Micela}, G., {et~al.} 2015{\natexlab{a}}, \aap,
  577, A132

\bibitem[{{Maldonado} {et~al.}(2015{\natexlab{b}}){Maldonado}, {Eiroa},
  {Villaver}, {Montesinos}, \& {Mora}}]{maldonado2015b}
{Maldonado}, J., {Eiroa}, C., {Villaver}, E., {Montesinos}, B., \& {Mora}, A.
  2015{\natexlab{b}}, \aap, 579, A20

\bibitem[{{Maldonado} {et~al.}(2018){Maldonado}, {Villaver}, \&
  {Eiroa}}]{maldonado2018}
{Maldonado}, J., {Villaver}, E., \& {Eiroa}, C. 2018, \aap, 612, A93

\bibitem[{{Mamajek} \& {Hillenbrand}(2008)}]{mamajek2008}
{Mamajek}, E.~E. \& {Hillenbrand}, L.~A. 2008, \apj, 687, 1264

\bibitem[{{Mayer} {et~al.}(2002){Mayer}, {Quinn}, {Wadsley}, \&
  {Stadel}}]{mayer2002}
{Mayer}, L., {Quinn}, T., {Wadsley}, J., \& {Stadel}, J. 2002, Science, 298,
  1756

\bibitem[{{Mayor} {et~al.}(2011){Mayor}, {Marmier}, {Lovis}, {Udry},
  {S{\'e}gransan}, {Pepe}, {Benz}, {Bertaux}, {Bouchy}, {Dumusque}, {Lo Curto},
  {Mordasini}, {Queloz}, \& {Santos}}]{mayor2011}
{Mayor}, M., {Marmier}, M., {Lovis}, C., {et~al.} 2011, ArXiv e-prints
  [\eprint[arXiv]{1109.2497}]

\bibitem[{{Mayor} {et~al.}(2003){Mayor}, {Pepe}, {Queloz}, {Bouchy},
  {Rupprecht}, {Lo Curto}, {Avila}, {Benz}, {Bertaux}, {Bonfils}, {Dall},
  {Dekker}, {Delabre}, {Eckert}, {Fleury}, {Gilliotte}, {Gojak}, {Guzman},
  {Kohler}, {Lizon}, {Longinotti}, {Lovis}, {Megevand}, {Pasquini}, {Reyes},
  {Sivan}, {Sosnowska}, {Soto}, {Udry}, {van Kesteren}, {Weber}, \&
  {Weilenmann}}]{mayor2003}
{Mayor}, M., {Pepe}, F., {Queloz}, D., {et~al.} 2003, The Messenger, 114, 20

\bibitem[{{Mints} \& {Hekker}(2017)}]{mints2017}
{Mints}, A. \& {Hekker}, S. 2017, \aap, 604, A108

\bibitem[{{Monet} {et~al.}(2003){Monet}, {Levine}, {Canzian}, {Ables}, {Bird},
  {Dahn}, {Guetter}, {Harris}, {Henden}, {Leggett}, {Levison}, {Luginbuhl},
  {Martini}, {Monet}, {Munn}, {Pier}, {Rhodes}, {Riepe}, {Sell}, {Stone},
  {Vrba}, {Walker}, {Westerhout}, {Brucato}, {Reid}, {Schoening}, {Hartley},
  {Read}, \& {Tritton}}]{monet2003}
{Monet}, D.~G., {Levine}, S.~E., {Canzian}, B., {et~al.} 2003, \aj, 125, 984

\bibitem[{{Mordasini} {et~al.}(2009){Mordasini}, {Alibert}, \&
  {Benz}}]{mordasini2009a}
{Mordasini}, C., {Alibert}, Y., \& {Benz}, W. 2009, \aap, 501, 1139

\bibitem[{{Mordasini} {et~al.}(2012){Mordasini}, {Alibert}, {Benz}, {Klahr}, \&
  {Henning}}]{mordasini2012}
{Mordasini}, C., {Alibert}, Y., {Benz}, W., {Klahr}, H., \& {Henning}, T. 2012,
  \aap, 541, A97

\bibitem[{{Mortier} {et~al.}(2013){Mortier}, {Santos}, {Sousa}, {Adibekyan},
  {Delgado Mena}, {Tsantaki}, {Israelian}, \& {Mayor}}]{mortier2013}
{Mortier}, A., {Santos}, N.~C., {Sousa}, S.~G., {et~al.} 2013, \aap, 557, A70

\bibitem[{{Mortier} {et~al.}(2012){Mortier}, {Santos}, {Sozzetti}, {Mayor},
  {Latham}, {Bonfils}, \& {Udry}}]{mortier2012}
{Mortier}, A., {Santos}, N.~C., {Sozzetti}, A., {et~al.} 2012, \aap, 543, A45

\bibitem[{{Mulders} {et~al.}(2016){Mulders}, {Pascucci}, {Apai}, {Frasca}, \&
  {Molenda-{\.Z}akowicz}}]{mulders2016}
{Mulders}, G.~D., {Pascucci}, I., {Apai}, D., {Frasca}, A., \&
  {Molenda-{\.Z}akowicz}, J. 2016, \aj, 152, 187

\bibitem[{{M{\"u}ller} {et~al.}(2013){M{\"u}ller}, {Roccatagliata}, {Henning},
  {Fedele}, {Pasquali}, {Caffau}, {Rodr{\'\i }guez-Ledesma}, {Mohler-Fischer},
  {Seemann}, \& {Klement}}]{muller2013}
{M{\"u}ller}, A., {Roccatagliata}, V., {Henning}, T., {et~al.} 2013, \aap, 556,
  A3

\bibitem[{{Nardetto} {et~al.}(2006){Nardetto}, {Mourard}, {Kervella},
  {Mathias}, {M{\'e}rand}, \& {Bersier}}]{nardetto2006}
{Nardetto}, N., {Mourard}, D., {Kervella}, P., {et~al.} 2006, \aap, 453, 309

\bibitem[{{Noyes} {et~al.}(1984){Noyes}, {Hartmann}, {Baliunas}, {Duncan}, \&
  {Vaughan}}]{noyes1984}
{Noyes}, R.~W., {Hartmann}, L.~W., {Baliunas}, S.~L., {Duncan}, D.~K., \&
  {Vaughan}, A.~H. 1984, \apj, 279, 763

\bibitem[{{Pollack} {et~al.}(1996){Pollack}, {Hubickyj}, {Bodenheimer},
  {Lissauer}, {Podolak}, \& {Greenzweig}}]{pollack1996}
{Pollack}, J.~B., {Hubickyj}, O., {Bodenheimer}, P., {et~al.} 1996, \icarus,
  124, 62

\bibitem[{{Robertson} {et~al.}(2012){Robertson}, {Endl}, {Cochran}, {MacQueen},
  {Wittenmyer}, {Horner}, {Brugamyer}, {Simon}, {Barnes}, \&
  {Caldwell}}]{robertson2012}
{Robertson}, P., {Endl}, M., {Cochran}, W.~D., {et~al.} 2012, \apj, 749, 39

\bibitem[{{Ryan}(1989)}]{ryan1989}
{Ryan}, S.~G. 1989, \aj, 98, 1693

\bibitem[{{Ryan} \& {Norris}(1991)}]{ryan1991}
{Ryan}, S.~G. \& {Norris}, J.~E. 1991, \aj, 101, 1835

\bibitem[{{Saar} {et~al.}(1998){Saar}, {Butler}, \& {Marcy}}]{saar1998}
{Saar}, S.~H., {Butler}, R.~P., \& {Marcy}, G.~W. 1998, \apj, 498, L153

\bibitem[{{Saar} \& {Donahue}(1997)}]{saar1997}
{Saar}, S.~H. \& {Donahue}, R.~A. 1997, \apj, 485, 319

\bibitem[{{Saar} \& {Fischer}(2000)}]{saar2000}
{Saar}, S.~H. \& {Fischer}, D. 2000, \apj, 534, L105

\bibitem[{{Santos} {et~al.}(2017){Santos}, {Adibekyan}, {Figueira},
  {Andreasen}, {Barros}, {Delgado-Mena}, {Demangeon}, {Faria}, {Oshagh},
  {Sousa}, {Viana}, \& {Ferreira}}]{santos2017}
{Santos}, N.~C., {Adibekyan}, V., {Figueira}, P., {et~al.} 2017, \aap, 603, A30

\bibitem[{{Santos} {et~al.}(2010{\natexlab{a}}){Santos}, {Gomes da Silva},
  {Lovis}, \& {Melo}}]{santos2010a}
{Santos}, N.~C., {Gomes da Silva}, J., {Lovis}, C., \& {Melo}, C.
  2010{\natexlab{a}}, \aap, 511, A54

\bibitem[{{Santos} {et~al.}(2001){Santos}, {Israelian}, \&
  {Mayor}}]{santos2001}
{Santos}, N.~C., {Israelian}, G., \& {Mayor}, M. 2001, \aap, 373, 1019

\bibitem[{{Santos} {et~al.}(2004){Santos}, {Israelian}, \&
  {Mayor}}]{santos2004}
{Santos}, N.~C., {Israelian}, G., \& {Mayor}, M. 2004, \aap, 415, 1153

\bibitem[{{Santos} {et~al.}(2010{\natexlab{b}}){Santos}, {Mayor}, {Benz},
  {Bouchy}, {Figueira}, {Lo Curto}, {Lovis}, {Melo}, {Moutou}, {Naef}, {Pepe},
  {Queloz}, {Sousa}, \& {Udry}}]{santos2010b}
{Santos}, N.~C., {Mayor}, M., {Benz}, W., {et~al.} 2010{\natexlab{b}}, \aap,
  512, A47

\bibitem[{{Santos} {et~al.}(2011){Santos}, {Mayor}, {Bonfils}, {Dumusque},
  {Bouchy}, {Figueira}, {Lovis}, {Melo}, {Pepe}, {Queloz}, {S{\'e}gransan},
  {Sousa}, \& {Udry}}]{santos2011}
{Santos}, N.~C., {Mayor}, M., {Bonfils}, X., {et~al.} 2011, \aap, 526, A112

\bibitem[{{Santos} {et~al.}(2007){Santos}, {Mayor}, {Bouchy}, {Pepe}, {Queloz},
  \& {Udry}}]{santos2007}
{Santos}, N.~C., {Mayor}, M., {Bouchy}, F., {et~al.} 2007, \aap, 474, 647

\bibitem[{{Santos} {et~al.}(2002){Santos}, {Mayor}, {Naef}, {Pepe}, {Queloz},
  {Udry}, {Burnet}, {Clausen}, {Helt}, {Olsen}, \& {Pritchard}}]{santos2002}
{Santos}, N.~C., {Mayor}, M., {Naef}, D., {et~al.} 2002, \aap, 392, 215

\bibitem[{{Santos} {et~al.}(2013){Santos}, {Sousa}, {Mortier}, {Neves},
  {Adibekyan}, {Tsantaki}, {Delgado Mena}, {Bonfils}, {Israelian}, {Mayor}, \&
  {Udry}}]{santos2013}
{Santos}, N.~C., {Sousa}, S.~G., {Mortier}, A., {et~al.} 2013, \aap, 556, A150

\bibitem[{{Smart} \& {Nicastro}(2014)}]{smart2014}
{Smart}, R.~L. \& {Nicastro}, L. 2014, \aap, 570, A87

\bibitem[{{Sneden}(1973)}]{sneden1973}
{Sneden}, C. 1973, \apj, 184, 839

\bibitem[{{Soubiran} \& {Girard}(2005)}]{soubiran2005}
{Soubiran}, C. \& {Girard}, P. 2005, \aap, 438, 139

\bibitem[{{Sousa} {et~al.}(2015){Sousa}, {Santos}, {Adibekyan}, {Delgado-
  Mena}, \& {Israelian}}]{sousa2015}
{Sousa}, S.~G., {Santos}, N.~C., {Adibekyan}, V., {Delgado- Mena}, E., \&
  {Israelian}, G. 2015, \aap, 577, A67

\bibitem[{{Sousa} {et~al.}(2011){Sousa}, {Santos}, {Israelian}, {Mayor}, \&
  {Udry}}]{sousa2011}
{Sousa}, S.~G., {Santos}, N.~C., {Israelian}, G., {Mayor}, M., \& {Udry}, S.
  2011, \aap, 533

\bibitem[{{Sousa} {et~al.}(2008){Sousa}, {Santos}, {Mayor}, {Udry},
  {Casagrande}, {Israelian}, {Pepe}, {Queloz}, \& {Monteiro}}]{sousa2008}
{Sousa}, S.~G., {Santos}, N.~C., {Mayor}, M., {et~al.} 2008, \aap, 487, 373

\bibitem[{{Sozzetti}(2004)}]{sozzetti2004}
{Sozzetti}, A. 2004, \mnras, 354, 1194

\bibitem[{{Sozzetti} {et~al.}(2007){Sozzetti}, {Torres}, {Charbonneau},
  {Latham}, {Holman}, {Winn}, {Laird}, \& {O'Donovan}}]{sozzetti2007}
{Sozzetti}, A., {Torres}, G., {Charbonneau}, D., {et~al.} 2007, \apj, 664, 1190

\bibitem[{{Sozzetti} {et~al.}(2006){Sozzetti}, {Torres}, {Latham}, {Carney},
  {Stefanik}, {Boss}, {Laird}, \& {Korzennik}}]{sozzetti2006}
{Sozzetti}, A., {Torres}, G., {Latham}, D.~W., {et~al.} 2006, \apj, 649, 428

\bibitem[{{Sozzetti} {et~al.}(2009){Sozzetti}, {Torres}, {Latham}, {Stefanik},
  {Korzennik}, {Boss}, {Carney}, \& {Laird}}]{sozzetti2009}
{Sozzetti}, A., {Torres}, G., {Latham}, D.~W., {et~al.} 2009, \apj, 697, 544

\bibitem[{{Tody}(1993)}]{tody1993}
{Tody}, D. 1993, in Astronomical Data Analysis Software and Systems II, ed.
  R.~J. {Hanisch}, R.~J.~V. {Brissenden}, \& J.~{Barnes}, Vol.~52, 173

\bibitem[{{Torres}(1999)}]{torres1999}
{Torres}, G. 1999, Publications of the Astronomical Society of the Pacific,
  111, 169

\bibitem[{{Torres} {et~al.}(2012){Torres}, {Fischer}, {Sozzetti}, {Buchhave},
  {Winn}, {Holman}, \& {Carter}}]{torres2012}
{Torres}, G., {Fischer}, D.~A., {Sozzetti}, A., {et~al.} 2012, \apj, 757, 161

\bibitem[{{Tsantaki} {et~al.}(2013){Tsantaki}, {Sousa}, {Adibekyan}, {Santos},
  {Mortier}, \& {Israelian}}]{tsantaki2013}
{Tsantaki}, M., {Sousa}, S.~G., {Adibekyan}, V.~Z., {et~al.} 2013, \aap, 555,
  A150

\bibitem[{{Udry} {et~al.}(2006){Udry}, {Mayor}, {Benz}, {Bertaux}, {Bouchy},
  {Lovis}, {Mordasini}, {Pepe}, {Queloz}, \& {Sivan}}]{udry2006}
{Udry}, S., {Mayor}, M., {Benz}, W., {et~al.} 2006, \aap, 447, 361

\bibitem[{{Yi} {et~al.}(2008){Yi}, {Kim}, {Demarque}, {Lee}, {Han}, \&
  {Kim}}]{yi2008}
{Yi}, S.~K., {Kim}, Y.-C., {Demarque}, P., {et~al.} 2008, in IAU Symposium,
  Vol. 252, The Art of Modeling Stars in the 21st Century, ed. L.~{Deng} \&
  K.~L. {Chan}, 413--416

\bibitem[{{Zechmeister} \& {K{\"u}rster}(2009)}]{zechmeister2009}
{Zechmeister}, M. \& {K{\"u}rster}, M. 2009, \aap, 496, 577

\end{thebibliography}

  \longtab{
	\begin{longtable}{l c c c c c c c}
	\caption{HARPS-N measurements for HD 220197 and HD 233832}\label{table:rv-table}\\
	\hline\hline
	BJD & T$_{exp}$ & RV  & BIS & \multicolumn{2}{c}{FWHM} & $\log R^{\prime}_{\rm HK}$ & Air mass\\
	    & [$\mathrm{s}$] & [$\mathrm{kms^{-1}}$] & [$\mathrm{kms^{-1}}$]&\multicolumn{2}{c}{[$\mathrm{kms^{-1}}$]} & & \\
	\hline
	\endfirsthead
	\caption{continued.}\\
	\hline\hline
	BJD & T$_{exp}$ & RV  & BIS & \multicolumn{2}{c}{FWHM} & $\log R^{\prime}_{\rm HK}$ & Air mass\\
	    & [$\mathrm{s}$] & [$\mathrm{kms^{-1}}$] & [$\mathrm{kms^{-1}}$]&\multicolumn{2}{c}{[$\mathrm{kms^{-1}}$]} & & \\
	\hline
	\endhead
	\hline
	\endfoot
	\multicolumn{8}{c}{HD 220197}\\
	\hline
		2456166.694961 & 900 & -40.2310 $\pm$ 0.0007 & -0.024 & 6.409 & (6.405) & -4.976 & 1.24\\
		2456174.590436 & 600 & -40.2275 $\pm$ 0.0006 & -0.024 & 6.400 & (6.395) & -4.985 & 1.03\\
		2456175.648876 & 600 & -40.2314 $\pm$ 0.0008 & -0.024 & 6.404 & (6.399) & -4.974 & 1.15\\
		2456180.609609 & 600 & -40.2322 $\pm$ 0.0007 & -0.029 & 6.411 & (6.405) & -4.967 & 1.08\\
		2456181.622981 & 600 & -40.2602 $\pm$ 0.0041 & -0.021 & 6.407 &  & -4.988 & 1.02\\
		2456198.602376 & 900 & -40.2287 $\pm$ 0.0014 & -0.022 & 6.386 & (6.377) & -4.774 & 1.22\\
		2456201.600188 & 600 & -40.2235 $\pm$ 0.0010 & -0.019 & 6.385 & (6.375) & -4.848 & 1.24\\
		2456295.310855 & 600 & -40.2369 $\pm$ 0.0007 & -0.023 & 6.440 & (6.413) & -4.974 & 1.13\\
		2456298.304272 & 600 & -40.2333 $\pm$ 0.0009 & -0.019 & 6.432 & (6.405) & -4.986 & 1.14\\
		2456299.306391 & 600 & -40.2356 $\pm$ 0.0007 & -0.020 & 6.443 & (6.415) & -4.968 & 1.16\\
		2456305.355553 & 600 & -40.2273 $\pm$ 0.0007 & -0.023 & 6.434 & (6.405) & -4.984 & 1.58\\
		2456322.303008 & 600 & -40.2312 $\pm$ 0.0010 & -0.017 & 6.336 & (6.304) & -4.926 & 1.53\\
		2456324.308237 & 600 & -40.2338 $\pm$ 0.0023 & -0.016 & 6.353 & (6.320) & -4.862 & 1.65\\
		2456324.321974 & 1256 & -40.2376 $\pm$ 0.0012 & -0.023 & 6.442 & (6.410) & -4.959 & 1.90\\
		2456483.729524 & 600 & -40.2306 $\pm$ 0.0008 & -0.020 & 6.458 & (6.391) & -4.970 & 1.02\\
		2456484.650796 & 600 & -40.2385 $\pm$ 0.0009 & -0.020 & 6.469 & (6.402) & -4.966 & 1.11\\
		2456485.727683 & 600 & -40.2362 $\pm$ 0.0007 & -0.020 & 6.467 & (6.401) & -4.964 & 1.02\\
		2456486.702419 & 600 & -40.2325 $\pm$ 0.0007 & -0.021 & 6.466 & (6.399) & -4.953 & 1.02\\
		2456487.721348 & 600 & -40.2358 $\pm$ 0.0010 & -0.020 & 6.475 & (6.408) & -4.931 & 1.02\\
		2456506.574807 & 600 & -40.2325 $\pm$ 0.0010 & -0.027 & 6.473 & (6.401) & -4.952 & 1.16\\
		2456530.649056 & 600 & -40.2395 $\pm$ 0.0007 & -0.025 & 6.481 & (6.404) & -4.963 & 1.07\\
		2456531.651444 & 600 & -40.2343 $\pm$ 0.0007 & -0.025 & 6.480 & (6.403) & -4.962 & 1.09\\
		2456532.735569 & 600 & -40.2390 $\pm$ 0.0007 & -0.027 & 6.483 & (6.405) & -4.953 & 1.53\\
		2456543.454059 & 600 & -40.2369 $\pm$ 0.0008 & -0.024 & 6.481 & (6.401) & -4.961 & 1.25\\
		2456544.661037 & 600 & -40.2350 $\pm$ 0.0007 & -0.023 & 6.474 & (6.394) & -4.968 & 1.24\\
		2456563.661916 & 600 & -40.2365 $\pm$ 0.0008 & -0.023 & 6.476 & (6.391) & -4.966 & 1.64\\
		2456565.619851 & 600 & -40.2381 $\pm$ 0.0006 & -0.022 & 6.484 & (6.399) & -4.971 & 1.33\\
		2456566.606976 & 600 & -40.2385 $\pm$ 0.0007 & -0.027 & 6.485 & (6.399) & -4.960 & 1.27\\
		2456579.602415 & 600 & -40.2355 $\pm$ 0.0007 & -0.017 & 6.480 & (6.391) & -4.976 & 1.49\\
		2456585.575694 & 600 & -40.2373 $\pm$ 0.0008 & -0.025 & 6.483 & (6.392) & -4.976 & 1.40\\
		2456602.563242 & 600 & -40.1661 $\pm$ 0.0004 & 0.026 & 6.352 &  & -5.000 & 1.76\\
		2456975.415640 & 600 & -40.2321 $\pm$ 0.0007 & -0.023 & 6.401 &  & -4.976 & 1.05\\
		2456998.337040 & 600 & -40.2360 $\pm$ 0.0008 & -0.024 & 6.404 &  & -4.984 & 1.03\\
		2456999.318083 & 600 & -40.2360 $\pm$ 0.0009 & -0.021 & 6.400 &  & -4.978 & 1.02\\
		2457002.320534 & 1200 & -40.2376 $\pm$ 0.0018 & -0.026 & 6.413 &  & -4.960 & 1.03\\
		2457003.323066 & 599 & -40.2442 $\pm$ 0.0010 & -0.020 & 6.404 &  & -4.990 & 1.03\\
		2457004.303645 & 600 & -40.2346 $\pm$ 0.0009 & -0.024 & 6.404 &  & -5.002 & 1.02\\
		2457023.326190 & 600 & -40.2368 $\pm$ 0.0014 & -0.028 & 6.405 &  & -5.023 & 1.16\\
		2457259.522203 & 600 & -40.2341 $\pm$ 0.0007 & -0.021 & 6.398 &  & -4.961 & 1.14\\
		2457269.670182 & 600 & -40.2347 $\pm$ 0.0008 & -0.022 & 6.402 &  & -4.946 & 1.21\\
		2457270.647244 & 600 & -40.2329 $\pm$ 0.0006 & -0.024 & 6.403 &  & -4.957 & 1.13\\
		2457271.679872 & 600 & -40.2338 $\pm$ 0.0007 & -0.023 & 6.404 &  & -4.960 & 1.29\\
		2457272.688865 & 600 & -40.2343 $\pm$ 0.0007 & -0.024 & 6.401 &  & -4.954 & 1.36\\
		2457273.623433 & 600 & -40.2325 $\pm$ 0.0008 & -0.018 & 6.402 &  & -4.958 & 1.09\\
		2457290.535927 & 600 & -40.2321 $\pm$ 0.0009 & -0.021 & 6.398 &  & -4.966 & 1.03\\
		2457291.504218 & 600 & -40.2307 $\pm$ 0.0007 & -0.022 & 6.399 &  & -4.970 & 1.02\\
		2457611.684273 & 600 & -40.2303 $\pm$ 0.0009 & -0.024 & 6.398 &  & -4.967 & 1.07\\
		2457623.530276 & 600 & -40.2265 $\pm$ 0.0008 & -0.017 & 6.391 &  & -4.986 & 1.12\\
		2457624.498685 & 600 & -40.2323 $\pm$ 0.0008 & -0.021 & 6.396 &  & -4.978 & 1.23\\
		2457644.645211 & 600 & -40.2359 $\pm$ 0.0008 & -0.021 & 6.401 &  & -4.962 & 1.22\\
		2457651.646254 & 600 & -40.2287 $\pm$ 0.0007 & -0.021 & 6.396 &  & -4.968 & 1.33\\
		2457652.647862 & 600 & -40.2330 $\pm$ 0.0009 & -0.023 & 6.397 &  & -4.986 & 1.35\\
		2457653.641366 & 600 & -40.2329 $\pm$ 0.0008 & -0.021 & 6.394 &  & -4.972 & 1.33\\
		2457654.644151 & 600 & -40.2352 $\pm$ 0.0011 & -0.025 & 6.406 &  & -4.990 & 1.37\\
		2457665.542415 & 600 & -40.2289 $\pm$ 0.0007 & -0.024 & 6.395 &  & -4.971 & 1.07\\
		2457679.563255 & 600 & -40.2288 $\pm$ 0.0008 & -0.027 & 6.401 &  & -4.961 & 1.29\\
		2457680.497913 & 600 & -40.2276 $\pm$ 0.0007 & -0.021 & 6.391 &  & -4.984 & 1.07\\
		2457681.537819 & 600 & -40.2286 $\pm$ 0.0007 & -0.019 & 6.394 &  & -4.990 & 1.19\\
		2457683.571934 & 600 & -40.2279 $\pm$ 0.0008 & -0.017 & 6.394 &  & -4.684 & 1.42\\
		2457701.573526 & 600 & -40.2304 $\pm$ 0.0018 & -0.029 & 6.409 &  & -4.970 & 2.11\\
		2457702.474375 & 600 & -40.2338 $\pm$ 0.0014 & -0.026 & 6.405 &  & -5.008 & 1.17\\
		2457727.401991 & 600 & -40.2281 $\pm$ 0.0007 & -0.021 & 6.395 &  & -5.000 & 1.17\\
		2457749.342299 & 600 & -40.2310 $\pm$ 0.0009 & -0.024 & 6.399 &  & -4.987 & 1.17\\
		2457750.357793 & 600 & -40.2353 $\pm$ 0.0013 & -0.024 & 6.394 &  & -5.002 & 1.26\\
		2457936.674180 & 600 & -40.2330 $\pm$ 0.0010 & -0.023 & 6.396 &  & -4.970 & 1.11\\
		2457937.674293 & 600 & -40.2289 $\pm$ 0.0009 & -0.021 & 6.396 &  & -4.981 & 1.10\\
		2457942.698513 & 600 & -40.2276 $\pm$ 0.0009 & -0.021 & 6.394 &  & -4.975 & 1.04\\
		2457943.725246 & 600 & -40.2297 $\pm$ 0.0008 & -0.017 & 6.391 &  & -4.975 & 1.02\\
		2457944.596109 & 600 & -40.2328 $\pm$ 0.0009 & -0.022 & 6.398 &  & -4.971 & 1.37\\
		2457952.703357 & 600 & -40.2354 $\pm$ 0.0011 & -0.020 & 6.399 &  & -4.984 & 1.02\\
		2457953.688599 & 600 & -40.2335 $\pm$ 0.0009 & -0.020 & 6.402 &  & -4.963 & 1.02\\
		2457954.694734 & 600 & -40.2341 $\pm$ 0.0008 & -0.022 & 6.401 &  & -4.971 & 1.02\\
		2457956.654577 & 600 & -40.2338 $\pm$ 0.0009 & -0.023 & 6.395 &  & -4.980 & 1.04\\
		2457971.637507 & 600 & -40.2334 $\pm$ 0.0008 & -0.021 & 6.395 &  & -4.975 & 1.02\\
		2457972.552032 & 600 & -40.2337 $\pm$ 0.0011 & -0.015 & 6.393 &  & -4.968 & 1.20\\
		2457973.660713 & 1800 & -40.2327 $\pm$ 0.0007 & -0.027 & 6.391 &  & -4.980 & 1.03\\
		2457974.592877 & 600 & -40.2330 $\pm$ 0.0010 & -0.018 & 6.400 &  & -4.969 & 1.07\\
		2457984.623067 & 600 & -40.2282 $\pm$ 0.0009 & -0.023 & 6.396 &  & -4.985 & 1.02\\
		2457989.541598 & 600 & -40.2311 $\pm$ 0.0008 & -0.018 & 6.401 &  & -4.976 & 1.09\\
		2457991.590650 & 600 & -40.2283 $\pm$ 0.0014 & -0.019 & 6.400 &  & -4.967 & 1.02\\
		2457993.596882 & 600 & -40.2289 $\pm$ 0.0008 & -0.021 & 6.397 &  & -4.963 & 1.02\\
		2457996.547027 & 600 & -40.2350 $\pm$ 0.0019 & -0.024 & 6.390 &  & -5.048 & 1.04\\
		2457997.502323 & 600 & -40.2284 $\pm$ 0.0012 & -0.021 & 6.395 &  & -4.972 & 1.14\\
		2458024.567725 & 600 & -40.2351 $\pm$ 0.0007 & -0.026 & 6.399 &  & -4.962 & 1.09\\
		2458026.634837 & 600 & -40.2332 $\pm$ 0.0014 & -0.021 & 6.399 &  & -4.976 & 1.44\\
		2458027.533479 & 1200 & -40.2336 $\pm$ 0.0019 & -0.022 & 6.401 &  & -4.972 & 1.05\\
		2458031.545700 & 600 & -40.2303 $\pm$ 0.0009 & -0.024 & 6.401 &  & -4.979 & 1.09\\
		2458044.476041 & 600 & -40.2338 $\pm$ 0.0008 & -0.016 & 6.396 &  & -4.965 & 1.03\\
	\hline
	\multicolumn{8}{c}{HD 233832}\\
	\hline
		2456344.717204 & 900 & -65.5142 $\pm$ 0.0007 & 0.007 & 5.735 & (5.660) & -4.934 & 1.25\\
		2456363.624524 & 900 & -65.5192 $\pm$ 0.0008 & 0.011 & 5.721 & (5.656) & -4.942 & 1.14\\
		2456364.633696 & 900 & -65.5173 $\pm$ 0.0009 & 0.006 & 5.720 & (5.655) & -4.963 & 1.17\\
		2456375.538791 & 900 & -65.5146 $\pm$ 0.0013 & 0.012 & 5.717 & (5.655) & -4.992 & 1.08\\
		2456376.484746 & 900 & -65.5150 $\pm$ 0.0009 & 0.008 & 5.722 & (5.659) & -4.992 & 1.09\\
		2456380.557851 & 900 & -65.5165 $\pm$ 0.0011 & 0.001 & 5.727 & (5.665) & -4.981 & 1.11\\
		2456381.614271 & 900 & -65.5162 $\pm$ 0.0008 & 0.006 & 5.721 & (5.659) & -4.980 & 1.25\\
		2456381.626365 & 900 & -65.5175 $\pm$ 0.0008 & 0.004 & 5.725 & (5.662) & -4.989 & 1.29\\
		2456405.551638 & 900 & -65.5135 $\pm$ 0.0015 & 0.013 & 5.722 & (5.654) & -4.957 & 1.26\\
		2456410.578812 & 900 & -65.5291 $\pm$ 0.0020 & -0.007 & 5.719 & (5.648) & -4.975 & 1.47\\
		2456424.522687 & 900 & -65.5144 $\pm$ 0.0009 & 0.007 & 5.741 & (5.660) & -5.011 & 1.37\\
		2457097.533774 & 900 & -65.4442 $\pm$ 0.0013 & -0.001 & 5.652 &  & -4.978 & 1.08\\
		2457099.561761 & 900 & -65.4451 $\pm$ 0.0012 & -0.005 & 5.647 &  & -4.988 & 1.08\\
		2457108.698979 & 900 & -65.4508 $\pm$ 0.0044 & 0.014 & 5.664 &  & -4.773 & 1.70\\
		2457109.471016 & 900 & -65.4425 $\pm$ 0.0012 & 0.009 & 5.665 &  & -4.972 & 1.10\\
		2457110.444016 & 900 & -65.4387 $\pm$ 0.0031 & 0.020 & 5.654 &  & -4.874 & 1.13\\
		2457112.440808 & 900 & -65.4433 $\pm$ 0.0030 & 0.016 & 5.666 &  & -4.841 & 1.12\\
		2457113.454520 & 900 & -65.4439 $\pm$ 0.0010 & 0.009 & 5.661 &  & -4.994 & 1.10\\
		2457114.407702 & 900 & -65.4425 $\pm$ 0.0020 & 0.008 & 5.661 &  & -4.929 & 1.18\\
		2457137.358615 & 900 & -65.4429 $\pm$ 0.0013 & 0.006 & 5.661 &  & -5.004 & 1.15\\
		2457142.478326 & 900 & -65.4430 $\pm$ 0.0011 & 0.002 & 5.665 &  & -4.992 & 1.12\\
		2457143.410021 & 900 & -65.4429 $\pm$ 0.0009 & 0.004 & 5.659 &  & -4.986 & 1.08\\
		2457144.439637 & 900 & -65.4410 $\pm$ 0.0009 & 0.010 & 5.668 &  & -4.974 & 1.08\\
		2457145.437333 & 900 & -65.4422 $\pm$ 0.0015 & 0.010 & 5.665 &  & -4.970 & 1.08\\
		2457154.430419 & 900 & -65.4378 $\pm$ 0.0023 & 0.006 & 5.653 &  & -4.987 & 1.10\\
		2457169.406042 & 900 & -65.4451 $\pm$ 0.0012 & 0.009 & 5.663 &  & -5.010 & 1.13\\
		2457170.403782 & 900 & -65.4431 $\pm$ 0.0009 & 0.006 & 5.659 &  & -4.986 & 1.13\\
		2457205.386928 & 900 & -65.4469 $\pm$ 0.0008 & 0.012 & 5.659 &  & -4.996 & 1.40\\
		2457209.382307 & 900 & -65.4488 $\pm$ 0.0017 & 0.013 & 5.663 &  & -4.966 & 1.44\\
		2457353.700256 & 1200 & -65.4538 $\pm$ 0.0021 & 0.002 & 5.662 &  & -4.945 & 1.36\\
		2457355.757480 & 1800 & -65.4618 $\pm$ 0.0027 & 0.012 & 5.673 &  & -4.976 & 1.14\\
		2457386.668107 & 900 & -65.4592 $\pm$ 0.0013 & 0.018 & 5.666 &  & -5.001 & 1.17\\
		2457387.723213 & 900 & -65.4528 $\pm$ 0.0013 & -0.003 & 5.656 &  & -4.988 & 1.08\\
		2457388.738421 & 900 & -65.4588 $\pm$ 0.0009 & 0.005 & 5.660 &  & -5.004 & 1.08\\
		2457389.719007 & 900 & -65.4570 $\pm$ 0.0013 & 0.010 & 5.656 &  & -4.973 & 1.08\\
		2457391.743411 & 900 & -65.4577 $\pm$ 0.0013 & 0.006 & 5.667 &  & -4.982 & 1.08\\
		2457417.741357 & 1800 & -65.4574 $\pm$ 0.0026 & 0.004 & 5.647 &  & -4.922 & 1.16\\
		2457418.738511 & 1800 & -65.4692 $\pm$ 0.0025 & 0.002 & 5.648 &  & -4.928 & 1.16\\
		2457421.757626 & 900 & -65.4616 $\pm$ 0.0025 & 0.019 & 5.658 &  & -4.932 & 1.22\\
		2457443.585531 & 900 & -65.4636 $\pm$ 0.0017 & 0.017 & 5.655 &  & -4.963 & 1.08\\
		2457444.579873 & 900 & -65.4694 $\pm$ 0.0034 & 0.007 & 5.669 &  & -5.014 & 1.08\\
		2457445.517548 & 900 & -65.4651 $\pm$ 0.0013 & 0.013 & 5.659 &  & -5.021 & 1.15\\
		2457472.514832 & 900 & -65.4662 $\pm$ 0.0012 & 0.012 & 5.663 &  & -5.042 & 1.07\\
		2457475.508171 & 900 & -65.4685 $\pm$ 0.0017 & 0.007 & 5.652 &  & -5.068 & 1.07\\
		2457491.397269 & 900 & -65.4672 $\pm$ 0.0009 & 0.009 & 5.659 &  & -5.025 & 1.13\\
		2457492.393470 & 900 & -65.4662 $\pm$ 0.0011 & 0.004 & 5.660 &  & -5.046 & 1.14\\
		2457501.482698 & 900 & -65.4685 $\pm$ 0.0010 & 0.003 & 5.654 &  & -5.029 & 1.10\\
		2457513.505807 & 900 & -65.4666 $\pm$ 0.0012 & 0.009 & 5.659 &  & -5.029 & 1.23\\
		2457521.469448 & 900 & -65.4629 $\pm$ 0.0010 & 0.014 & 5.658 &  & -5.045 & 1.19\\
		2457522.384150 & 900 & -65.4690 $\pm$ 0.0013 & 0.009 & 5.662 &  & -5.043 & 1.08\\
		2457523.426551 & 900 & -65.4722 $\pm$ 0.0033 & -0.001 & 5.659 &  & -5.149 & 1.11\\
		2457525.399486 & 900 & -65.4712 $\pm$ 0.0015 & 0.004 & 5.652 &  & -5.088 & 1.09\\
		2457526.405188 & 900 & -65.4683 $\pm$ 0.0011 & 0.010 & 5.663 &  & -5.058 & 1.10\\
		2457540.378563 & 900 & -65.4718 $\pm$ 0.0011 & 0.017 & 5.664 &  & -5.032 & 1.11\\
		2457552.387606 & 900 & -65.4725 $\pm$ 0.0021 & 0.006 & 5.662 &  & -5.096 & 1.20\\
		2457553.396578 & 900 & -65.4720 $\pm$ 0.0015 & 0.007 & 5.650 &  & -5.042 & 1.23\\
		2457573.391404 & 900 & -65.4727 $\pm$ 0.0009 & 0.006 & 5.658 &  & -5.043 & 1.47\\
		2457574.382765 & 900 & -65.4727 $\pm$ 0.0013 & 0.008 & 5.649 &  & -5.063 & 1.44\\
		2457703.767152 & 900 & -65.4826 $\pm$ 0.0012 & 0.003 & 5.660 &  & -5.041 & 1.26\\
		2457763.587833 & 900 & -65.4860 $\pm$ 0.0015 & 0.000 & 5.655 &  & -5.067 & 1.33\\
		2457773.685240 & 900 & -65.4816 $\pm$ 0.0013 & 0.008 & 5.657 &  & -5.033 & 1.08\\
		2457790.737442 & 900 & -65.4862 $\pm$ 0.0010 & 0.005 & 5.659 &  & -5.031 & 1.19\\
		2457808.567046 & 900 & -65.4873 $\pm$ 0.0010 & 0.013 & 5.657 &  & -5.058 & 1.09\\
		2457810.564248 & 900 & -65.4883 $\pm$ 0.0028 & 0.004 & 5.666 &  & -5.115 & 1.08\\
		2457812.620392 & 900 & -65.4854 $\pm$ 0.0010 & 0.005 & 5.653 &  & -5.027 & 1.09\\
		2457833.458789 & 900 & -65.4896 $\pm$ 0.0017 & 0.005 & 5.655 &  & -5.006 & 1.14\\
		2457852.540479 & 900 & -65.4910 $\pm$ 0.0016 & 0.010 & 5.656 &  & -4.992 & 1.13\\
		2457853.541938 & 900 & -65.4907 $\pm$ 0.0022 & 0.010 & 5.657 &  & -5.103 & 1.14\\
		2457862.448992 & 900 & -65.4888 $\pm$ 0.0010 & 0.013 & 5.666 &  & -5.057 & 1.07\\
		2457895.461297 & 900 & -65.4914 $\pm$ 0.0010 & 0.011 & 5.656 &  & -5.035 & 1.23\\
		2457932.386154 & 900 & -65.4927 $\pm$ 0.0041 & 0.002 & 5.665 &  & -5.009 & 1.35\\
		2457933.398122 & 900 & -65.4976 $\pm$ 0.0035 & 0.009 & 5.662 &  & -5.012 & 1.42\\
		2457934.397580 & 900 & -65.4947 $\pm$ 0.0041 & 0.011 & 5.661 &  & -5.004 & 1.44\\
		2457935.393821 & 900 & -65.4943 $\pm$ 0.0061 & 0.004 & 5.656 &  & -5.021 & 1.43\\
		2458047.760373 & 900 & -65.4990 $\pm$ 0.0025 & 0.004 & 5.656 &  & -5.114 & 1.61\\
		2458075.754804 & 900 & -65.5039 $\pm$ 0.0012 & 0.013 & 5.655 &  & -5.132 & 1.24\\
		2458189.733233 & 900 & -65.5064 $\pm$ 0.0011 & 0.005 & 5.652 &  & -5.114 & 1.64\\
		2458190.690636 & 900 & -65.5087 $\pm$ 0.0010 & 0.004 & 5.659 &  & -5.118 & 1.37\\
		2458268.439049 & 1800 & -65.5074 $\pm$ 0.0018 & 0.005 & 5.666 &  & -5.107 & 1.25\\
		2458269.473627 & 900 & -65.5076 $\pm$ 0.0009 & 0.005 & 5.664 &  & -5.050 & 1.39\\
	\hline
	\end{longtable}
	}
	
\end{document}